\documentclass{aa}

\usepackage{caption}
\usepackage{subfig}
\usepackage{graphicx}
\usepackage{natbib}
\usepackage{txfonts}

\begin{document}
 \title{Determining Energy Balance in the Flaring Chromosphere from Oxygen V Line Ratios}
 \titlerunning{Oxygen V Footpoint Diagnostics}

 
 \author{D. R. Graham\thanks{current address: INAF-Osservatorio Astrofisico di Arcetri, I-50125 Firenze, Italy}, L. Fletcher, and N. Labrosse}
 
 \offprints{D. R. Graham}

 \institute{SUPA School of Physics and Astronomy, University of Glasgow, Glasgow G12 8QQ, U.K.\\
 \email{dgraham@arcetri.astro.it}
 }
 

 \date{Received ; accepted}

 
 \abstract
  {The impulsive phase of solar flares is a time of rapid energy deposition and heating in the lower solar atmosphere, leading to changes in the temperature and density structure of the region.}
  {We use an \ion{O}{V} density diagnostic formed of the $\lambda 192/\lambda 248$ line ratio, provided by the {\it Hinode}/EIS instrument, to determine the density of flare footpoint plasma at \ion{O}{V} formation temperatures of $\sim 2.5\times 10^5$K, giving a constraint on the properties of the heated transition region. }
  {{\it Hinode}/EIS rasters from 2 small flare events in December 2007 were used. Raster images were co-aligned to identify and establish the footpoint pixels, multiple-component Gaussian line fitting of the spectra was carried out to isolate the density diagnostic pair, and the density was calculated for several footpoint areas. The assumptions of equilibrium ionisation and optically-thin radiation for the \ion{O}{V} lines used were assessed and found to be acceptable. Properties of the electron distribution, for one of the events, were deduced from earlier RHESSI hard X-ray observations and used to calculate the plasma heating rate, within 2 semi-empirical atmospheres, delivered by an electron beam adopting collisional thick-target assumptions. The radiative loss rate for this plasma was also calculated for comparison with possible energy input mechanisms.}
  {Electron number densities of up to $10^{11.9}~{\rm cm^{-3}}$ were measured during the flare impulsive phase using the \ion{O}{V} $\lambda 192/ \lambda248$ diagnostic ratio. The heating rate delivered by an electron beam was found to exceed the radiative losses at this density, corresponding to a height of 450~km, and when assuming a completely ionised target atmosphere far exceed the losses but at a height of 1450-1600~km. A chromospheric thickness of 70-700~km was found to be required to balance a conductive input to the \ion{O}{V}-emitting region with radiative losses.}
  {Electron densities have been observed in footpoint sources, at transition region temperatures, comparable with previous results but with improved spatial information. The observed densities can be explained by heating of the chromosphere by collisional electrons, with \ion{O}{V} formed at heights of 450-1600 km above the photosphere, depending on the atmospheric ionisation fraction.}
  

   \maketitle
%

\section{Introduction}\label{sec:intro}
Solar flares are the result of a sudden reconfiguration of the active region magnetic field, releasing on the order of $10^{31}$~ergs of energy (for a medium-sized event) in a matter of minutes. They are visible across the electromagnetic spectrum, but particularly in the dramatic appearance of high temperature extreme ultraviolet (EUV) and soft X-ray emitting plasmas. The maximum intensity from the EUV and SXR plasmas  occur roughly coincident with, or slightly lagging,  an abrupt hard X-ray (HXR) burst emitted by electrons that have been accelerated to non-thermal energies and lose energy via bremsstrahlung radiation in a dense, collisional plasma. HXR imaging spectroscopy by the Reuven Ramaty High Energy Solar Spectroscopic Imager (RHESSI) \citep{2002SoPh..210....3L} has successfully shown that this emission is mainly confined to compact \emph{footpoint} regions in the low solar atmosphere --- chromosphere to low-corona --- and in some cases dense loop sources. The energy collisionally deposited in the chromosphere by electron-electron and electron-ion collisions is capable of explaining rapid heating of footpoint plasma, to millions of kelvin. Theoretical models of collisional energy deposition and the response of the atmosphere have a long heritage and are now very detailed in terms of their treatment of dynamics and radiation transfer \citep[e.g.][]{1984ApJ...279..896N,2005ApJ...630..573A,2009ApJ...702.1553L}; observational constraints are of course needed to test whether the models are adequate.


How these electrons are accelerated out of an initially (almost) thermal distribution remains a mystery, and decades of effort have focused on using observational signatures to try and understand this. This is not straightforward:  observational signatures, other than those directly produced by the accelerated electrons themselves (i.e. HXR and gyrosynchrotron radiation) tend to say more about the heated medium than about the energetic particles that heat it, though direct collisional effects from non-thermal particles may be detectable from detailed examination of spectral features \citep[e.g.][]{2011A&A...533A..81K}. However, the distribution of density and temperature in the flare-heated atmosphere, as a function of time, bears some relation to the distribution in space and time of the heating, and thus to the distribution of non-thermal particles which are thought responsible for the collisional heating. The standard `collisional thick target' model for the heating of the flare atmosphere invokes 
collisional loss by non-thermal electrons (remote from their acceleration location) and efforts to model the flare atmospheric response have tended to focus on this.  

Theoretical problems with this model, primarily the large beam number density required by observations and the associated difficulties with the beam propagating stably, have led to alternatives being proposed to the collisional thick target model, such as local acceleration or re-acceleration of electrons in the chromosphere \citep{2008ApJ...675.1645F,2009A&A...508..993B}, while flare heating of the chromosphere by wave dissipation \cite{2013ApJ...765...81R} or by thermal conduction \citep{2013ApJ...767...83G} have also been suggested. None of these proposed mechanisms have been implemented into atmospheric modeling so far, or their radiation signatures evaluated.

Flare energy deposition leads to rapid heating and ionisation of compact regions of the lower solar atmosphere to nearly 10~MK \citep[e.g.][]{2004A&A...415..377M,2013ApJ...771..104F,2013ApJ...767...83G} and expansion of the local plasma, known as chromospheric evaporation, observed spectroscopically in the EUV and SXR \citep[e.g.][]{1983SoPh...86...67A,1999ApJ...521L..75C,2009ApJ...699..968M}. The resulting flaring solar atmosphere is strongly stratified in density and temperature, and we need observations to completely specify its properties over a broad range of temperatures. Interpreting the atmospheric structure from optically thin observations of footpoint radiation is extremely challenging as height information can not be directly extracted. We can however measure the density at specific temperatures and compare these to the structure of a modelled pre-flare or flaring atmospheres. In this paper we focus on the properties of plasma at a temperature of $2.5\times 10^5$K ($\log T = 5.4$) using a density sensitive O~{\sc v} ratio $\lambda 192/ \lambda248$.


Previous work with \emph{Hinode}/EIS has identified flare footpoint densities of a few times $10^{10}$ to $10^{11}{\rm cm^{-3}}$ \citep{2010ApJ...719..213W,2011A&A...526A...1D,2011A&A...532A..27G} at temperatures of 1-2~MK, but higher densities were documented in the late 70's and early 80's using the \emph{Skylab} NRL normal incidence slit spectrograph S082B \citep{1977ApOpt..16..879B} and the Ultraviolet Spectrometer and Polarimeter (UVSP) instrument on board the Solar Maximum Mission (SMM) \citep{1980SoPh...65...73W}. \cite{1977ApJ...215..329D} and \cite{1977ApJ...215..652F} derived flare densities from O~{\sc iv} ratios at $10^{5}~{\rm K}$ from \emph{Skylab} slit data. The spectral profiles had distinct velocity-shifted components with densities of $10^{11}$ and $10^{12}~{\rm cm^{-3}}$ in the stationary, and at least $10^{13}~{\rm cm^{-3}}$ in the downward moving component. At similar line formation temperatures \cite{1982ApJ...253..353C} used the O~{\sc iv} 1401\AA\ to Si~{\sc iv} 1402.7\AA\ ratio and found that a pre-flare density of $2.5 \times 10^{11}~{\rm cm^{-3}}$ rose to $3 \times 10^{12}~{\rm cm^{-3}}$ during the flare impulsive phase. This enhancement was also located within a footpoint kernel and coincided temporally with a HXR burst. The limited spatial resolution in these instruments did however make the distinction between loop and footpoint sources difficult.

In this paper we make new measurements of flare footpoint densities at transition region temperatures finding similar enhanced densities, although addressing the earlier ambiguity in identifying the emitting region by employing the capabilities offered by \emph{Hinode}/EIS. Section 2 introduces the data and spectral fitting of the diagnostic lines, Section 3 reveals the measured footpoint densities, Section 4 considers the validity of the assumption of optically thin plasma in ionisation equilibrium, and finally Sections 5 \& 6 interpret the results within the framework of an atmosphere heated collisionally by electrons, and one heated solely by a conductive flux.


\section{Hinode/EIS Data}

\begin{figure}
  \centering
  \includegraphics[width=7cm]{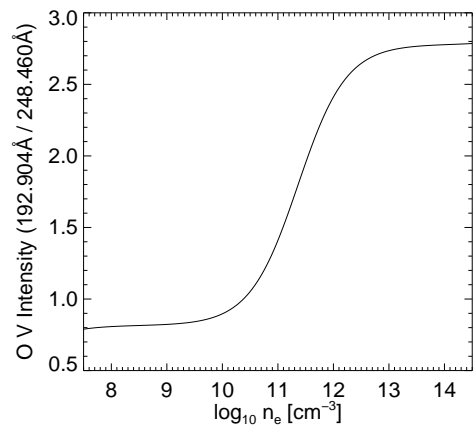}
  \caption{Diagnostic ratio for O {\sc v} 192.904\AA\ / 248.460\AA\ }
  \label{fig:ov_ratio}%
\end{figure}

Good diagnostics of density below $\log~T$ = 6.0 are uncommon in the EIS spectral range and, as far as we are aware, have not been used in the study of flare footpoint plasmas. Transition region lines during the flare impulsive phase are observed to rise almost simultaneously with the HXR response \citep{1984ApJ...280..457P, 1985SoPh...96..317M} and density measurements of the emitting plasma at these temperatures are therefore extremely useful to help locate the heated plasma within the atmosphere in relation to the HXR emission. The O {\sc v} $n = 3$ transitions at 192.904\AA\ and 248.460\AA\ are observed by \emph{Hinode}/EIS and their ratio forms a diagnostic pair sensitive to densities up to around $10^{12}~{\rm cm^{-3}}$ \citep{1982ApJ...257..913W}. A formation temperature at equilibrium of 250,000~K places their maximum emission within the transition region in the quiet-sun \citep{1981ApJS...45..635V} and the diagnostic curve is shown in Figure \ref{fig:ov_ratio}.

In this article we analyse data from these lines during the impulsive phases of two C-class flares; SOL2007-12-14T01:39 (Flare 1) and SOL2007-12-14T14:16 (Flare 2). The flares have been the subject of previous papers; in an emission measure analysis \citep{2013ApJ...767...83G}, and evaporation and non-thermal broadening study \citep{2009ApJ...699..968M, 2011ApJ...740...70M} respectively.

The impulsive phase in each flare was mostly captured in one raster scan by EIS and the raster closest to the GOES SXR derivative peak was chosen to best represent the time of maximum energy deposition \citep[see][]{2013ApJ...767...83G}. RHESSI coverage was not available for Flare 1, but the presence of non-thermal RHESSI HXR footpoints was confirmed for Flare 2 \citep{2009ApJ...699..968M}. Footpoints were identified in both flares by the sudden appearance of compact enhancements in the cooler EIS lines, e.g He {\sc ii} and Fe {\sc viii} at $\log$ T = 4.5 - 5.7, explained by a rise in the chromospheric temperature, and enhanced electron densities in the Fe {\sc xii} - {\sc xiv} diagnostics at temperatures between $\log$~T = 6.1 - 6.3 \citep{2011ApJ...740...70M}, again a sign that the dense lower atmosphere is being heated \citep{2011A&A...532A..27G}. Again we refer the reader to the previous work in \cite{2013ApJ...767...83G} for more details on the raster selection and footpoint 
identification.

\begin{figure*}[t]
  \centering
   \includegraphics[width=14cm]{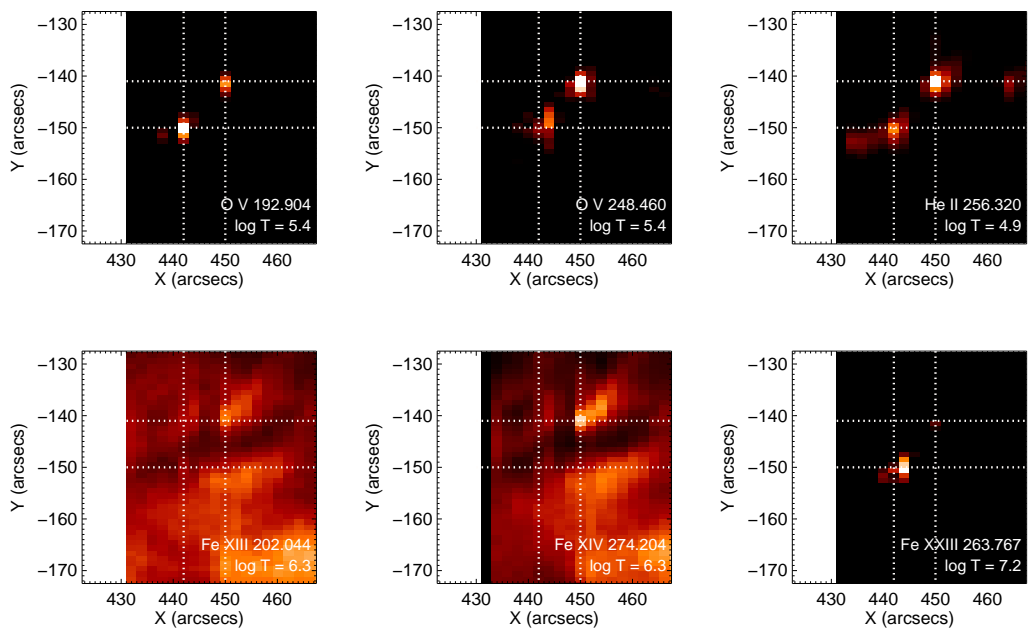}
   \caption{Fitted intensities in several different temperature lines for the 01:35:28~UT raster (Flare 1) after correcting for the EIS detector alignment. White dotted lines show the position of the footpoints in each image relative to the O {\sc v}~192\AA\ line. The x-offset is highlighted by the black strip on the left of the Fe~{\sc xiv} 274\AA\ image.}
  \label{fig:crosscorr}%
\end{figure*}

Both flares used the same CAM\_ARTB\_RHESSI\_b\_2 study having an exposure time of 10 sec using the 2\arcsec\ slit covering a 40\arcsec\ x 144\arcsec\ area. The slit stepping for these rasters was continuous, not sparse, in the $x$ direction, advantageous in these smaller events, and crossed the footpoints during the impulsive phase. The rasters were prepared with the standard SolarSoft {\sc eis\_prep} routine.

Time and wavelength dependent changes in the absolute calibration of both the EIS spectrograph detectors were recently re-characterised by \cite{2013A&A...555A..47D} and \cite{2014ApJS..213...11W}, finding that both detectors are degrading at independent rates; this complicates the O~{\sc v} diagnostic as the lines fall on separate detectors (see next section). The data in our study is from relatively early in the \emph{Hinode} mission, yet the new calibrations vary significantly from the pre-flight set. For our analysis we use the \cite{2014ApJS..213...11W} calibration by applying the {\sc eis\_recalibrate\_intensity.pro} routine.

\subsection{EIS Image Alignment}\label{sec:wavecorr}

The diagnostic lines in question occur on the two different CCDs of the EIS spectrograph, which are not identically aligned with respect to the grating. This introduces a fixed spatial offset in the slit direction (y-axis) between features observed on each CCD. In addition, the short wavelength CCD is tilted slightly relative to the grating, creating a further wavelength-dependent offset (longer wavelengths appearing lower on the CCD). In the case of a diagnostic ratio it is extremely important to remove these effects to ensure that we observe the same feature in both wavelengths. The tilt was characterised by \cite{2009A&A...495..587Y} and a correction included in the {\sc eis\_ccd\_offset.pro} routine which returns the y-offset required to correct both the grating tilt and spatial offset.

We also performed our own cross-correlation using the Fe~{\sc xiii} 202\AA\ and Fe~{\sc xiv} 274\AA\ lines for a range of binary thresholds. We found an average shift of 16.7\arcsec\ in the y-axis (slit axis) but also a shift of 1 pixel, or 2.0\arcsec, in the x-axis. Our y-offset was comparable with the 17.06\arcsec\ shift given by the {\sc eis\_ccd\_offset} routine, affirming the \cite{2009A&A...495..587Y} prediction. The additional 2\arcsec\ x-offset was also documented by \cite{2007PASJ...59S.727Y} and later attributed to an optics focusing effect; mostly corrected for in 24 August 2008. The only complication with alignment in the x-direction is that the same feature is seen in two consecutive slit steps, and correcting for it effectively halves the observing cadence.


In Figure \ref{fig:crosscorr} we show example fitted intensity images in several emission lines from the Flare 1 raster after correlation. Here we use our measured x-shift while using the wavelength dependent y-offset calculated by the EIS routine. The white dotted lines cross the same (x, y) location in each image, centred on the footpoints in O {\sc v} 192\AA, and intersect the corresponding bright emission in all lines for the northern footpoint. The southern footpoint has an anomalous x-shift of around 2\arcsec\ when observed at some wavelengths, which is curious as the shift is not apparent in its northern companion, ruling out a problem with the correlation. It is also not temperature sensitive, as both Fe {\sc xxiii} and O {\sc v} are equally offset while the He~{\sc ii}~256\AA\ image shows no change. The offset is puzzling as an instrumental effect would be expected to alter the other footpoint's position, yet there is no obvious relation to the line temperature or wavelength. We can not discount the effects of optical depth here, as the active region lies at 450\arcsec\ on the solar disc with a 29 degree projection to Earth. A small filament seen in TRACE images (not shown here) lies just to the north of the footpoint, perhaps absorbing some emission from lines with a higher opacity. We discuss the opacity in Section \ref{sec:assumptions}.


\subsection{Line Fitting Technique}

Multiple component Gaussian fits with several constraints are required to extract the diagnostic line intensities. The oxygen 248\AA\ line is relatively unblended but the density-sensitive 192\AA\ transition lies in a challenging part of the EIS spectral range. Several oxygen and iron lines formed at different temperatures lie in close proximity and interpreting their individual intensities is difficult, or sometimes impossible without constraints from other lines in the raster. Significant contributions, as predicted by the CHIANTI v7.1.3 atomic database \citep{1997A&AS..125..149D,2013ApJ...763...86L}, are listed in Table \ref{tab:lines}.\\

\begin{figure}
  \centering
  \includegraphics[width=7cm]{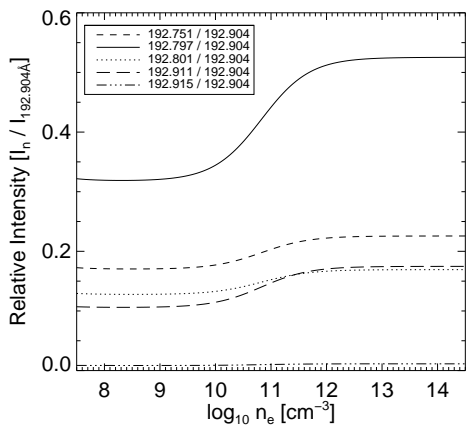}
  \caption{Density sensitivity of the blended O~{\sc v} lines with respect to the 192.904\AA\ transition.}
  \label{fig:192_ratios}%
\end{figure}

\noindent {\bf 192\AA} - The target 192.904\AA\ line lies in a complex profile of six O {\sc v} lines, two coronal Fe {\sc xi} lines, and a hot Ca {\sc xvii} line; including a linear background makes a total of 29 free parameters. In order to reduce the required number of free parameters we use a fitting routine based on the technique described in \cite{2009ApJ...697.1956K}. The neighbouring five O {\sc v} lines are first constrained by fixing their intensities, centroids, and line widths relative to the target 192.904\AA\ fit. The intensity ratios are estimated from CHIANTI for an electron density of $10^{12}~{\rm cm^{-3}}$ and are shown in Table \ref{tab:lines}. There is some density sensitivity among the other oxygen lines relative to 192.904\AA, plotted in Figure \ref{fig:192_ratios}, however only the 192.797\AA\ line has any significant density sensitivity, and only up to $10^{12}~{\rm cm^{-3}}$. As we do not know the density before fitting we have tested that the choice of density will have a minimal 
effect. Changing the density from $10^{12}$ to $10^{10}~{\rm cm^{-3}}$ raises the fitted 192.904\AA\ intensity by $<10\%$. By using the higher density of $10^{12}~{\rm cm^{-3}}$ it serves to reduce the diagnostic ratio, erring on the side of a lower density estimate.

Fe {\sc xi} 192.627\AA\ is distinct from the overall profile and can be fitted with little constraint, only requiring that the width be $\pm 40\%$ of the well-observed Fe~{\sc xi} 188.230\AA\ line. Fe~{\sc xi} 192.814\AA\ is tied to the Fe {\sc xi} 188.230\AA\ line by a constant intensity ratio of 0.2, which is insensitive to density. The centroid shift of both iron lines within the profile were also tied to the 188\AA\ line. Finally, the parameters of the last remaining Ca {\sc xvii} line are left mostly unconstrained, with the only limit imposed being that the centroid position should not be red-shifted.\\


\noindent {\bf 248\AA} - The 248\AA\ O {\sc v} line is more straightforward to fit as the Ar {\sc xiii} contribution is in most cases resolvable. Al {\sc viii} however, lies near the O~{\sc v} line centroid and some extra constraint is needed. From CHIANTI, using the DEM derived in \cite{2013ApJ...767...83G}, the Al~{\sc viii} contribution is $10\%$ of the O {\sc v} intensity for photospheric abundances.


\begin{table}
\caption{Emission lines within the 192\AA\ and 248\AA\ profiles. The O~{\sc v} 192\AA\ intensities relative to the 192.904\AA\ line are taken from CHIANTI v7.1.3. for a density of $n_e~=~10^{12}~{\rm cm^{-3}}$.} 
\label{tab:lines}
\begin{center}
\begin{tabular}{l c c c}
\hline
Ion & Wavelength (\AA) & $\log {\rm T_{max}} (K)$ & Relative Intensity\\ \hline
O {\sc v} & 192.751 & 5.4 & 0.22\\
O {\sc v} & 192.797 & 5.4 & 0.51\\
O {\sc v} & 192.801 & 5.4 & 0.17\\
O {\sc v} & 192.904 & 5.4 & 1.0\\
O {\sc v} & 192.911 & 5.4 & 0.17\\
O {\sc v} & 192.915 & 5.4 & 0.011\\
Fe {\sc xi} & 192.627 & 6.2 & -\\
Fe {\sc xi} & 192.814 & 6.2 & -\\
Ca {\sc xvii} & 192.858 & 6.8 & - \\

\hline
O {\sc v}     & 248.461 & 5.4 & -\\
Al {\sc viii} & 248.459 & 5.9 & -\\
Ar {\sc xiii} & 248.697 & 6.4 & -\\

\hline
\end{tabular}
\end{center}
\end{table}

\subsection{Footpoint Pixel Selection}

\begin{figure*}
  \centering
   \subfloat{{(a)}\includegraphics[width=11cm]{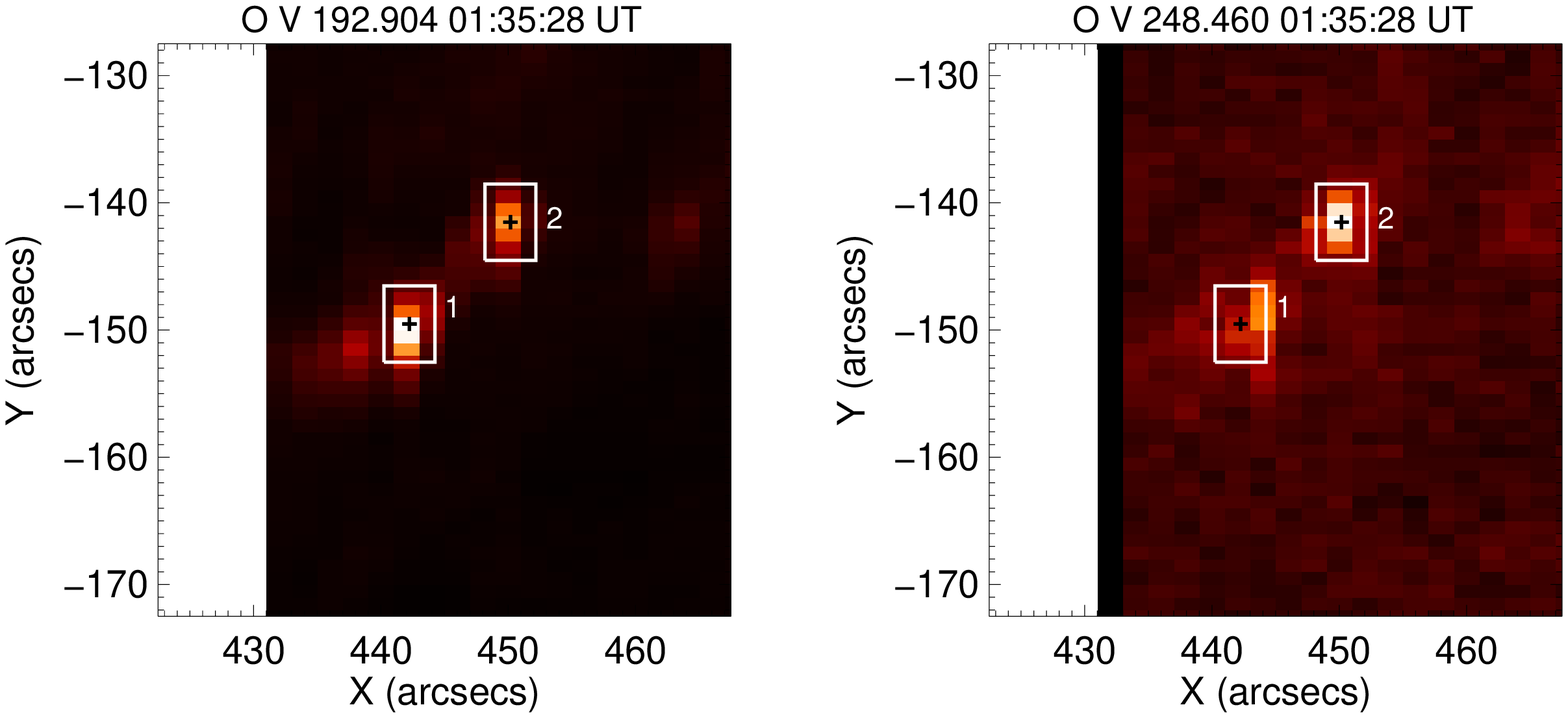}}
   
   \subfloat{{(b)}\includegraphics[width=11cm]{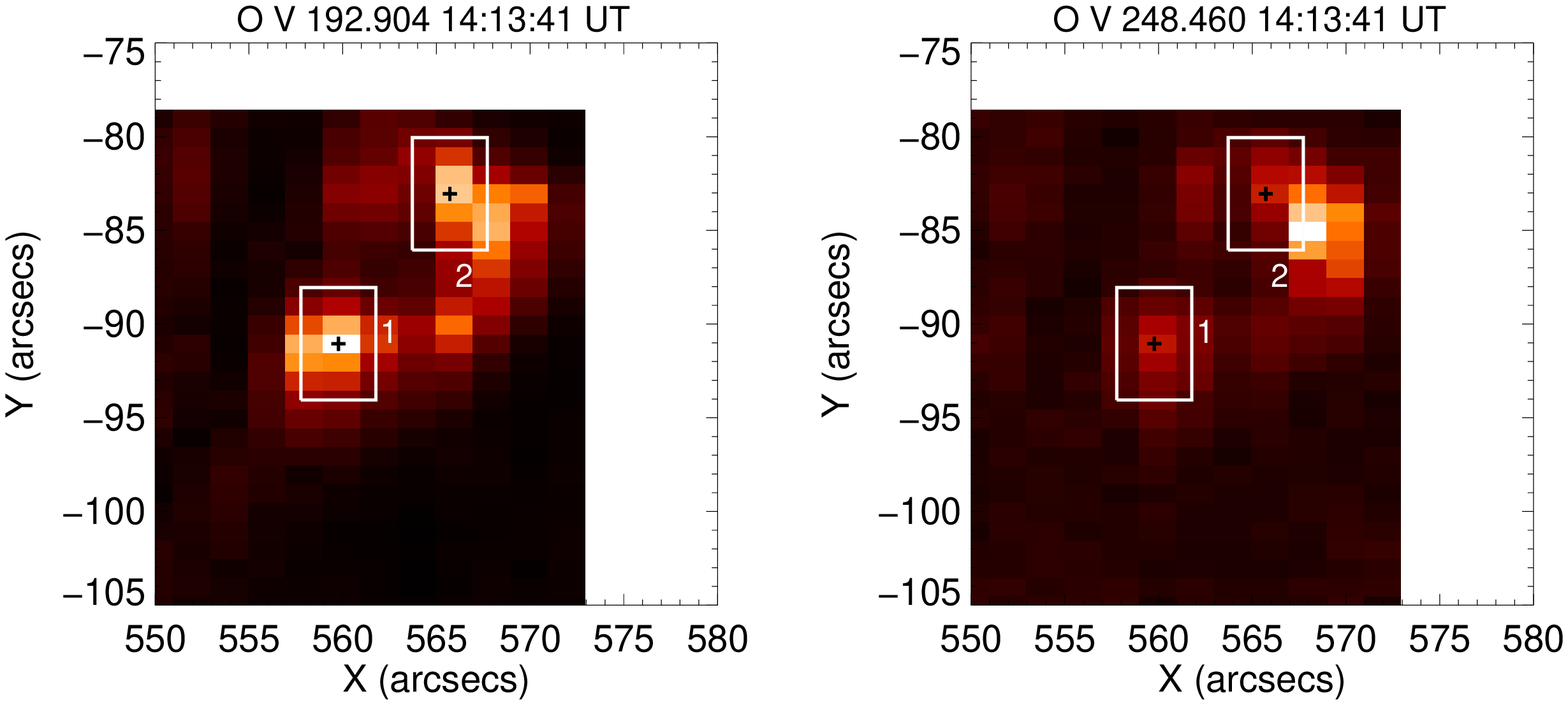}}
   \caption{Intensity maps for both 192.904\AA\ and 248.460\AA\ lines corresponding to the impulsive phase for both flares. Two footpoints regions (1 and 2) are bounded in each flare by a white box (pixels under the line are included in the selection). The brightest pixel within each footpoint region is also identified by a black cross.}
  \label{fig:footmap}%
\end{figure*}

In Figure \ref{fig:footmap} we show the intensity maps in the 192.904\AA\ and 248.460\AA\ lines (left and right columns) for each flare (rows (a) and (b)). The compact footpoints are clearly identified against the background active region emission. In each flare two footpoint regions are defined by 6\arcsec\ x 7\arcsec\ white boxes; including all of the pixels within and under the white lines. The brightest 192\AA\ pixel in each footpoint region is marked by a black cross on both wavelengths. We noted that the brightest emission in both 192\AA\ and 248\AA\ lines did not always coincide. This is particularly clear in Flare 1 (Figure \ref{fig:footmap} (a)), where for Footpoint 1 the strongest emission in 248\AA\ is found one pixel (2\arcsec) to the right of 192\AA\ (discussed in the previous section).


To acknowledge these small position uncertainties the intensity ratio is calculated using two methods. In the first method, spectra for each footpoint are averaged over the region within the white box, which is the same for both lines, before fitting and finding the ratio. The second method takes the ratio from the pixel brightest in 192\AA\ within the footpoint regions and makes the ratio with same pixel in 248\AA; plus a binning by $\pm$1 pixels in the y-direction (a 2\arcsec x 3\arcsec region) to account for any correlation uncertainty. Densities from the box method are likely to return a lower limit on the density as it includes pixels from outside of the footpoint regions. The latter `pixel' method assumes that the brightest emission originates from the same source in both lines but has enough spread to account for uncertainties in the correlation. Of course if the source size is smaller than the instrumental point spread function (3-4\arcsec in EIS) then the density in unresolved structures may yet be higher.


\section{Results}

\subsection{Fitted Spectra}

\begin{figure*}
  \centering
  \subfloat{{(a)}\includegraphics[width=13cm]{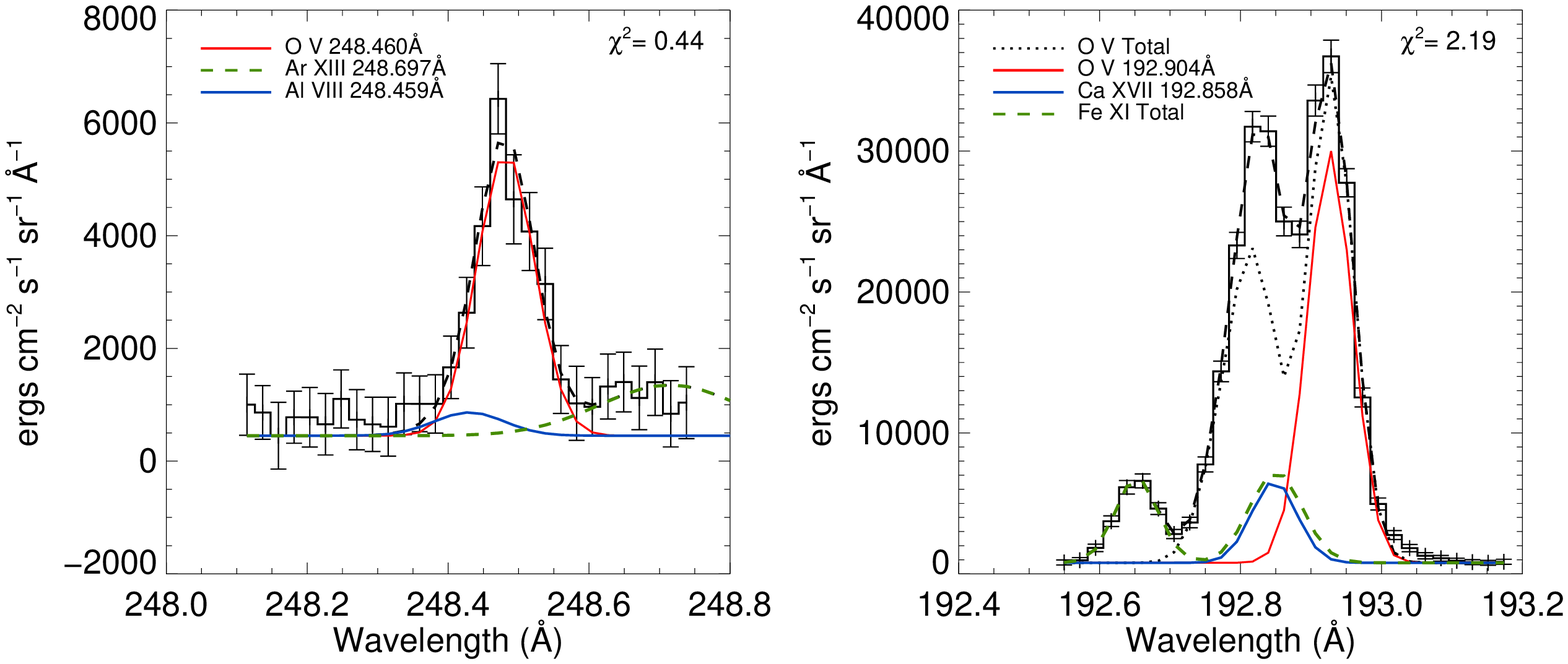}}
  
  \subfloat{{(b)}\includegraphics[width=13cm]{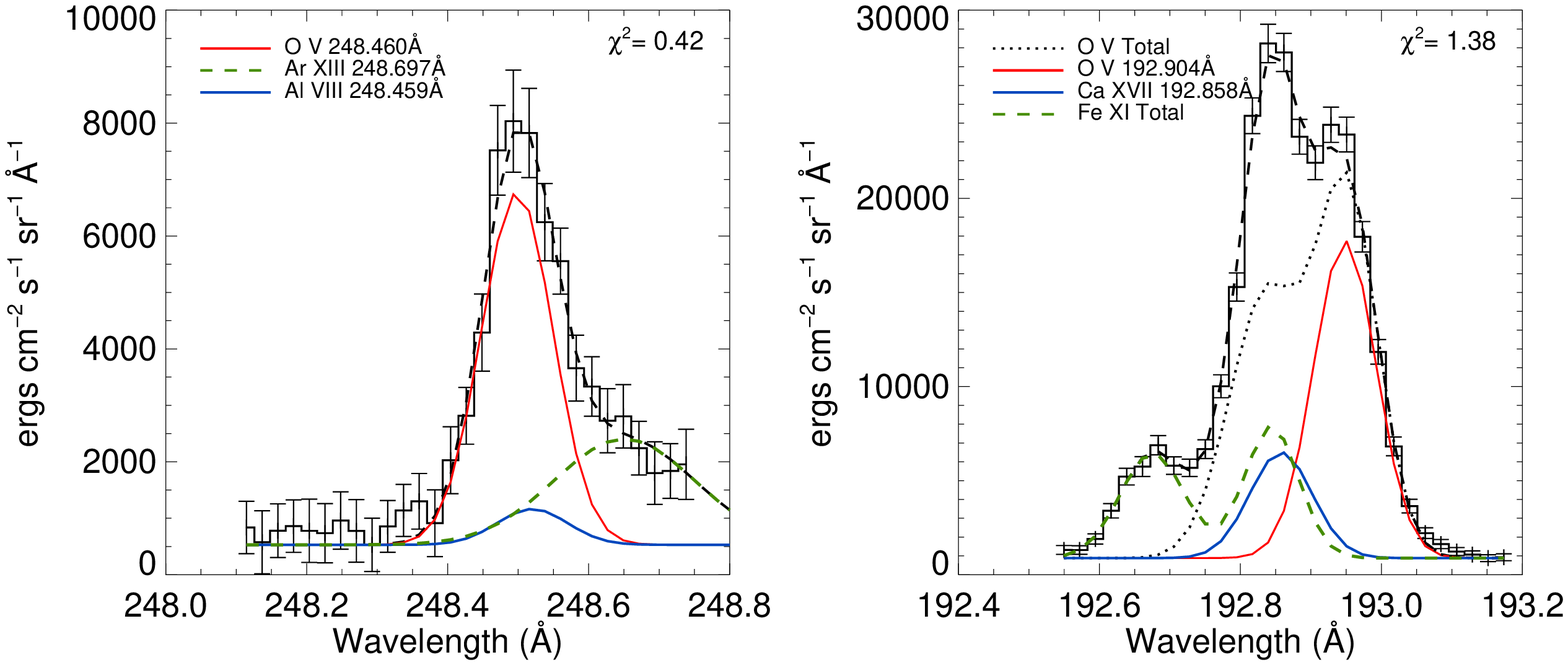}}
  \caption{Fitted spectra for Flare 1 using the brightest pixels for Footpoints 1 and 2 - plots (a) and (b) respectively. For both spectral profiles the total fit and data is shown in black solid and dashed lines respectively. The calibrated 1-$\sigma$ errors on each data point are also plotted including the reduced $\chi^2$ value for the fit. The oxygen lines forming the diagnostic ratio are plotted in red in both columns, with the various blends described by the legend.}
  \label{fig:spectra1}%
\end{figure*}

\begin{figure*}
  \centering
  \subfloat{{(a)}\includegraphics[width=13cm]{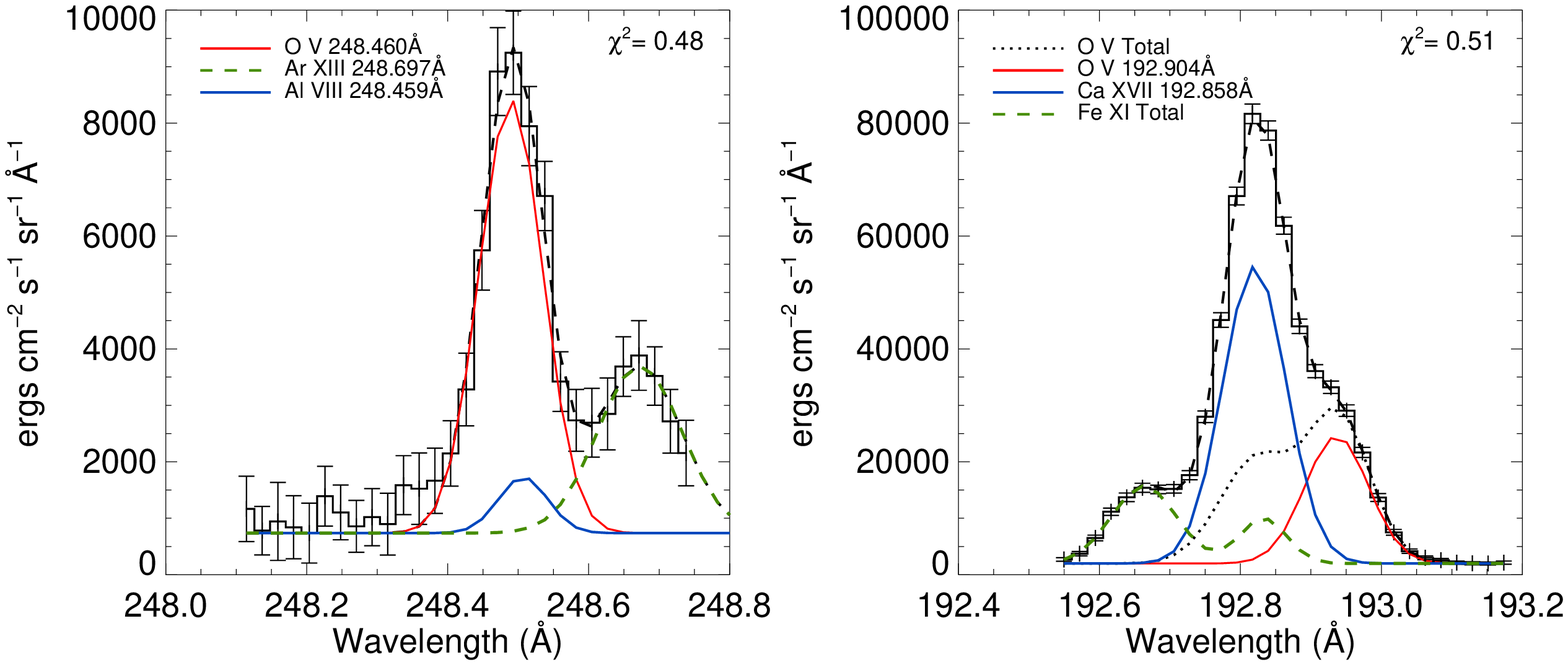}}
  
  \subfloat{{(b)}\includegraphics[width=13cm]{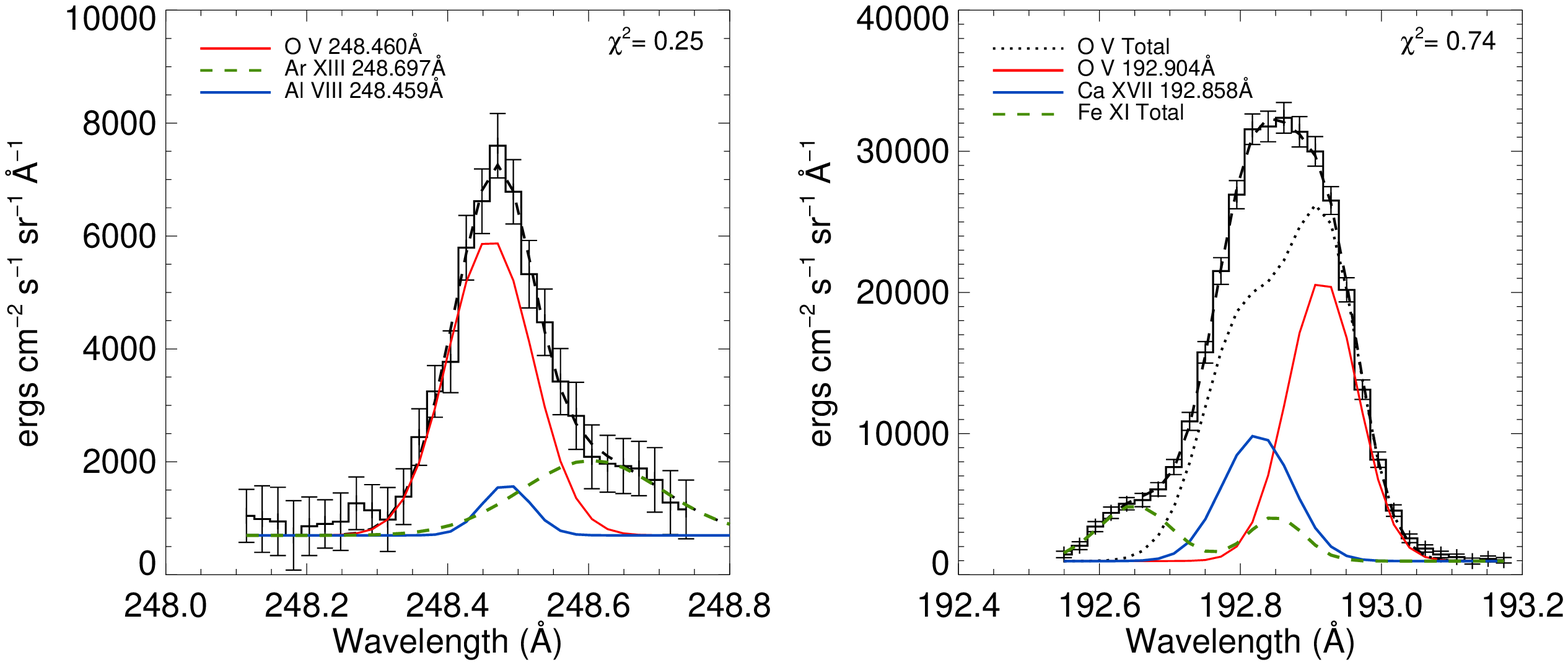}}
  \caption{As for Figure \ref{fig:spectra1} but showing spectra from Flare 2, Footpoints 1 and 2 - plots (a) and (b) respectively}
  \label{fig:spectra2}%
\end{figure*}

\begin{figure*}
  \centering
  \includegraphics[width=13cm]{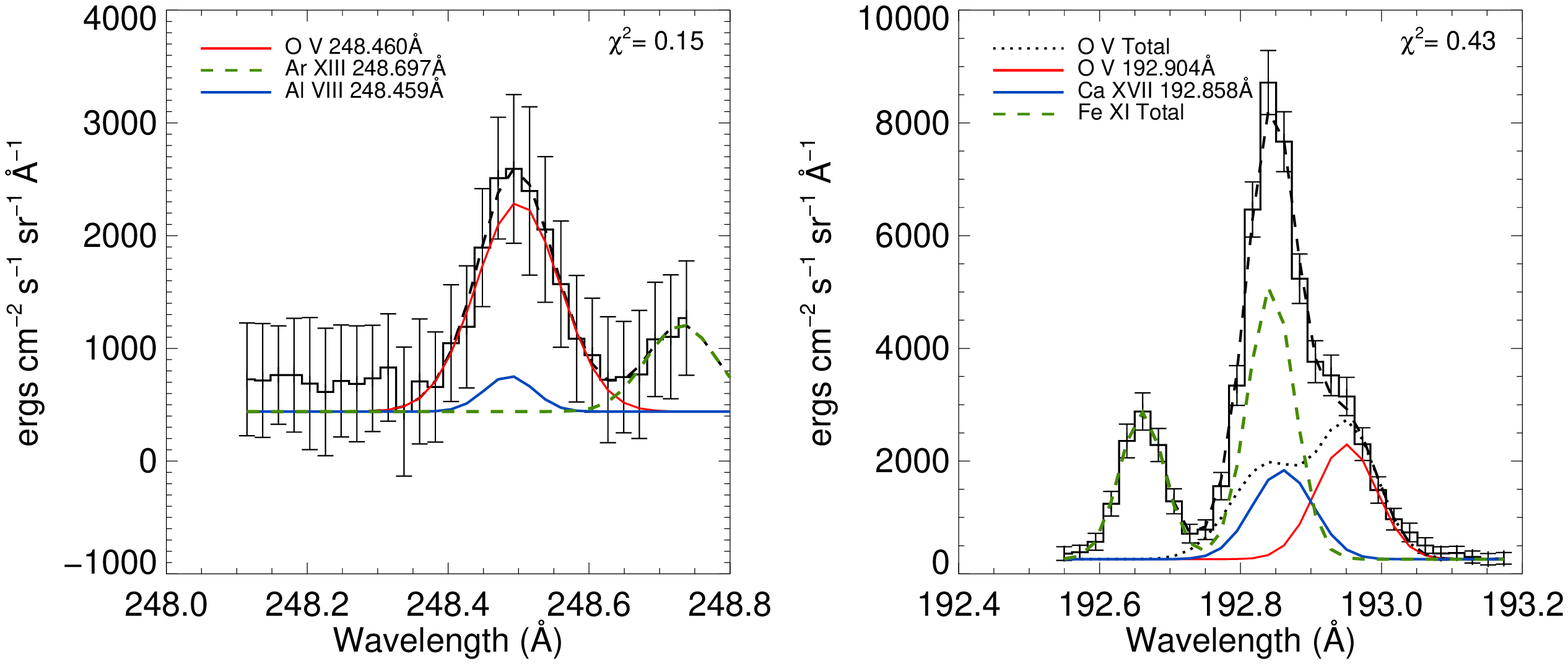}
  \caption{As described in Figure \ref{fig:spectra1} but with a sample spectra taken from a non-flaring region 55\arcsec\ to the north of the Flare 1 footpoints - note the dominant iron Fe~{\sc xi} contribution in the green dashed line.}
  \label{fig:spectraback}%
\end{figure*}

The fitted spectra for the bright pixels in Figure \ref{fig:spectra1} (Flare 1) show a clear enhancement of the O~{\sc v} 192\AA\ line (red line, right hand profiles) compared to the pre-flare (Figure \ref{fig:spectraback}) with the line being distinct from the rest of the profile. This 192.904\AA\ transition is excited from a metastable level and is expected to brighten quickly as the dense emitting plasma is heated; similar line profiles in transition region brightenings were also found by \cite{2007PASJ...59S.727Y}. The unconstrained Ca~{\sc xvii} emission (blue line) in these footpoints makes a small contribution among the other predicted Fe~{\sc xi} and O~{\sc v} intensities (green dashed and black dotted lines respectively). In Figure \ref{fig:spectra2} the Ca~{\sc xvii} (blue line) emission is more significant, possibly due to more hot loop material being present along the line of sight. 

We have highlighted the spectra partly to demonstrate the robustness of the fitting method. Between the constrained intensities of the Fe~{\sc xi} and O~{\sc v} lines there is a small range of parameter space that the free O~{\sc v}~192.904 and Ca~{\sc xvii}~192.853 lines can occupy. The spectra here demonstrate that large changes in the intensity ratio between the calcium and oxygen lines are handled well among the various constraints, for example, nowhere do we see predicted line intensities for Fe~{\sc xi}~188.213 or the five constrained O~{\sc v} lines exceeding the profile boundary. Assuming that the atomic physics used to predict the constraints is to be trusted, these two free lines can be extracted from the blend reliably.


The O~{\sc v} 248\AA\ fits (left hand plot on Figures \ref{fig:spectra1} and \ref{fig:spectra2}) are more straightforward, however, the EIS effective area is lower here and the data uncertainties are larger. The Al~{\sc viii} contribution (blue line) in the red-wing can be fitted with minimal constraint. The background level is noisy but the line here is distinct enough to be removed from the O~{\sc v} 248\AA\ line. Ar~{\sc xiii} (green dashed line) lies within the O~{\sc v} profile and was fitted with a fixed 10\% contribution of the O~{\sc v} line. Accounting for these blends will reduce the 248\AA\ line intensity; as the line is the denominator overestimating the blends will increase the density estimate.

\subsection{Densities}

The footpoint densities were found by comparing the measured ratio to the diagnostic curve in Figure \ref{fig:ov_ratio}, and our results are given in Table \ref{tab:densitiescal}. Taking the absolute error from the fit parameters for width and peak intensity, the error in total intensity of the 192\AA\ and 248\AA\ lines can be calculated, and thus the error on the ratio. An estimate of the density error is made by taking these limits on the ratio and finding an upper and lower limit to the density on the diagnostic curve.

In all of the footpoints we find that the `pixel' derived density is between $\log n_e~({\rm cm^{-3}}) = 11.3 - 11.9$, in line with the measurements from \emph{Skylab} and UVSP where the density varied between $\log n_e = 11.5 - 12.5$ \citep{1982ApJ...253..353C, 1991ApJ...382..349K}. Flare 2 FP2 suggests that the density may be slightly over $\log n_e = 12.0$ given the uncertainties and our conservative fitting.

Checking the results for the box-averaged method reveals similar densities with a tendency to lower values in all pixels. In Section \ref{sec:wavecorr} we noted that Flare 1 FP1 has an unexplained shift in the position of the footpoint in 248\AA\ relative to 192\AA\ and was omitted from the pixel-method results. Using the box method it shows a density of $\log n_e = 11.34$. The EIS pixel size is most likely large compared to the EUV footpoint area and this is some evidence to show that higher densities may be revealed when observing on smaller scales; indeed recent results from the Interface Region Imaging Spectrograph \citep[IRIS][]{2014SoPh..289.2733D} are beginning to show that the footpoint scales are below the EIS spatial resolution \citep[for example][]{2015ApJ...799..218Y}.





\begin{table*}
\caption{O {\sc v} footpoint densities derived from the $I_{192}/I_{248}$ ratio. Density estimates denoted by an asterisk have an upper limit beyond the maximum of the diagnostic curve; the density therefore either represents a lower limit or has no estimated upper limit. The pixel values use a source of 2\arcsec\ x 3\arcsec\ and the box is taken from a 6\arcsec\ x 7\arcsec\ region.}
\begin{center}
\begin{tabular}{l c c c c}
\hline
Footpoint & $I_{192}/I_{248}$ & $\log {\rm n_e}$ & $I_{192}/I_{248}$ & $\log {\rm n_e}$ \\ 
          & (pixel) & $({\rm cm^{-3}})$ & (box) & $({\rm cm^{-3}})$  \\
\hline
Flare 1 FP1 & -                         & -                            & 1.78 $\pm 0.14$ & 11.34 $\pm 0.25$ \\
Flare 1 FP2 & 1.74 $\pm 0.12$ & 11.31 $\pm 0.20$ & 1.09 $\pm 0.04$ & 10.57 $\pm 0.17$ \\
Flare 2 FP1 & 2.17 $\pm 0.20$ & 11.71$\pm 0.43$  & 2.08 $\pm 0.05$ & 11.62 $\pm 0.10$ \\
Flare 2 FP2 & 2.33 $\pm 0.07$ & 11.89 $\pm 0.18$ & 1.32 $\pm 0.03$ & 10.91 $\pm 0.07$ \\

\hline
\label{tab:densitiescal}
\end{tabular}
\end{center}
\end{table*}

%


\section{Understanding the Assumptions}\label{sec:assumptions}

The high densities in Flare 2 should be examined carefully considering the effect of relaxing the assumptions of optically thin plasma in ionisation equilibrium, under which the CHIANTI diagnostic curves are calculated. In this section we calculate the optical depth for both diagnostic lines in two semi-empirical atmospheric models: the VAL-E bright network model \citep{1981ApJS...45..635V}, representing the case that the atmosphere is heated without having time to respond hydrodynamically, and the F1 model \citep{1980ApJ...242..336M} representing the structure of an atmosphere that has already adapted to flare energy input. Additionally, we investigate the diagnostic curves for non-equilibrium line formation temperatures.

\subsection{Optical Depth}


The optical depth at line centre for both O~{\sc v} lines is calculated using the expression in \cite{1978stat.book.....M} (used to the same effect in \cite{2011A&A...531A.122D} for Si~{\sc iii} transitions). Assuming a thermal Gaussian absorption profile, where $\Delta \nu_D = (\nu_0 / c) (2kT / M)^{1/2}$, the optical depth at line centre is given by

\begin{equation}\label{eq:tau}
 \tau_{\nu = \nu_o} = {B_{ij} \over {4 \pi^{3/2}}} {h \nu \over \Delta \nu_D} \int \left( 1 - {g_i n_j \over g_j n_i} \right) {n_i \over n_{ion}} {n_{ion} \over n_{el}} {n_{el} \over n_H} {n_H \over n_e} n_e dS,
\end{equation}

where $B_{ij}$ is the Einstein absorption coefficient from the lower level $i$ to upper level $j$. $B_{ij}$ can be expressed in terms of the radiative decay coefficient, 

\begin{equation}
 B_{ij} = A_{ji} \left( {g_j \over g_i} \right) \left( {c^2 \over 2 h \nu^3} \right),
\end{equation}

where $g_i$ and $g_j$ are the statistical weights for levels $i$ and $j$ and $A_{ji}$ is the Einstein coefficient for spontaneous de-excitation. Crucially, the terms inside the integral depend on position in the atmosphere; the relative population of level $i$, $n_i/n_{ion}$ is sensitive to density, the ion abundance $n_{ion} / n_{el}$ is strongly dependent on temperature, and $n_H/n_e$ varies with depth. These parameters are taken from the CHIANTI database and we assume constant photospheric abundances, $n_{el}/n_H$, given by \cite{1998SSRv...85..161G}.

\begin{figure}
  \centering
   \subfloat{{(a)}\includegraphics[width=9cm]{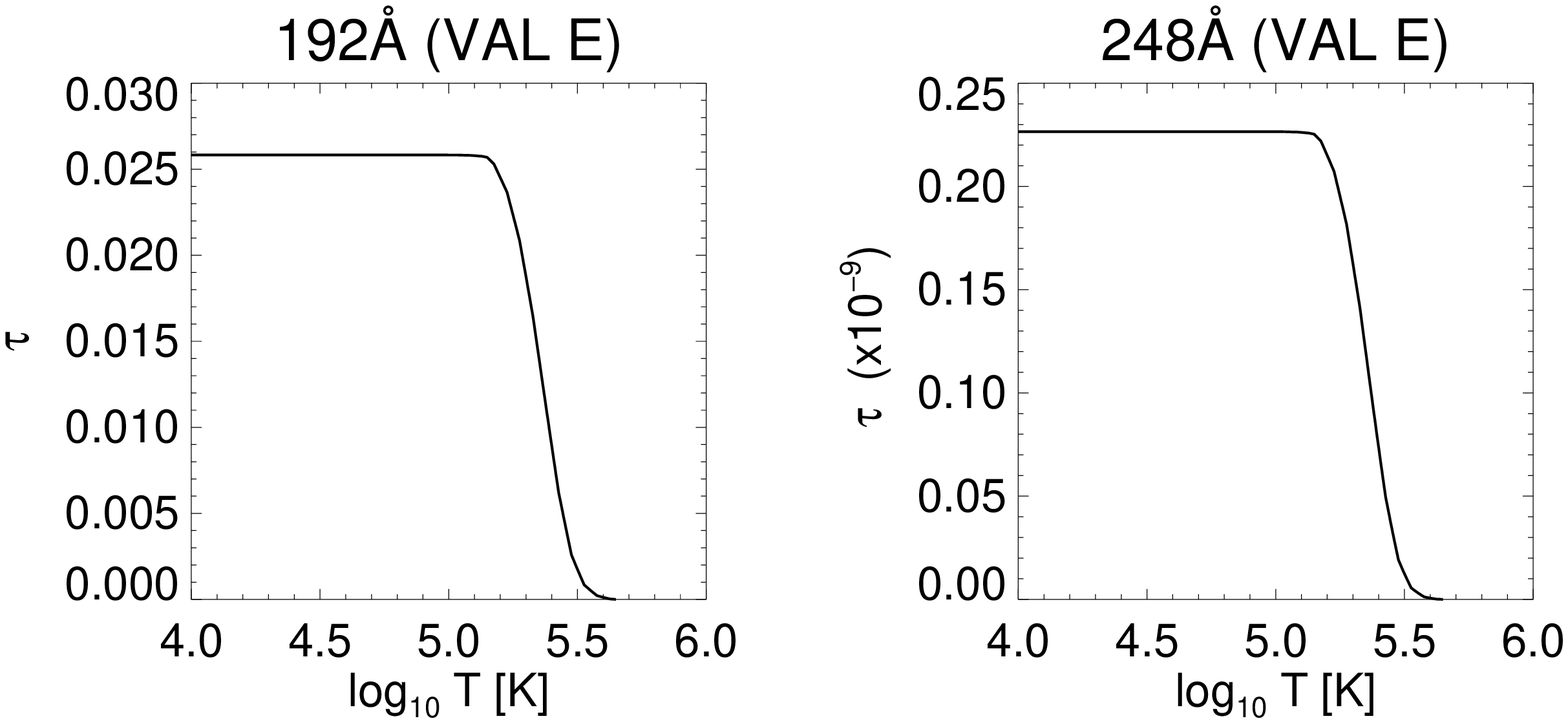}}

   \subfloat{{(b)}\includegraphics[width=9cm]{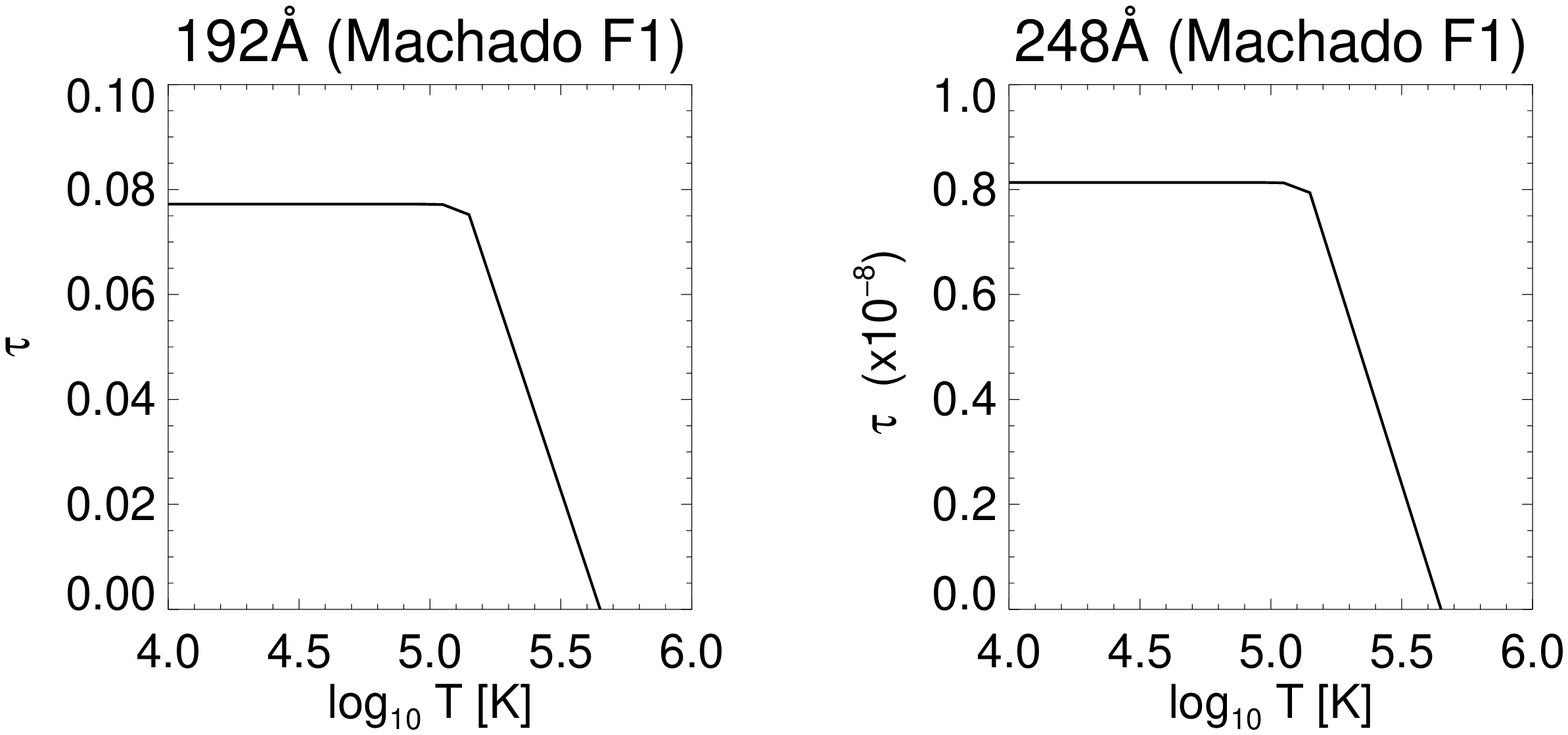}}
   \caption{Optical depth at line centre for the oxygen lines in both the VAL-E bright network (top row) and F1 flare atmospheres (bottom row) as function of temperature.}
  \label{fig:ov_opacity}%
\end{figure}

The optical depth at line centre is found for both lines by integrating from a chosen depth through the atmosphere above it, and in Figures \ref{fig:ov_opacity} (a) and (b) we show how it varies with temperature for both the VAL-E and the F1 atmospheres. The optical depth quickly plateaus where the ion abundance peaks, as the absorbing ion is not yet formed at lower temperatures. The 192\AA\ line always remains below $\tau = 1$ whereas 248\AA\ remains completely optically thin with $\tau < 10^{-8}$. The large difference in optical depths between the lines is mostly due to the difference in population of their lower levels. The optically thin assumption in this case is valid for both lines, and the diagnostic ratio should not be altered.

We should however note that Equation \ref{eq:tau} considers only the balance between the absorption of photons and spontaneous and stimulated emission, where the absorption profile is equal to the emission profile. A more complete description would include the redistribution of the level population within the ion due to the difference in opacity between the lines, or if the absorption and emission profiles are not equal. Certainly the analytical work by \cite{2004ApJ...613L.181K} and observations by \cite{2014ApJ...784L..39K} have shown that viewing-angle dependent enhancements, or suppressions, over the expected optically thin ratio are possible if one line has a significant optical depth. Without performing a similar detailed calculation it is not clear which way our diagnostic ratio would be altered, if at all.


\subsection{Ionisation Equilibrium}

\begin{figure}
  \centering
  \includegraphics[width=8cm]{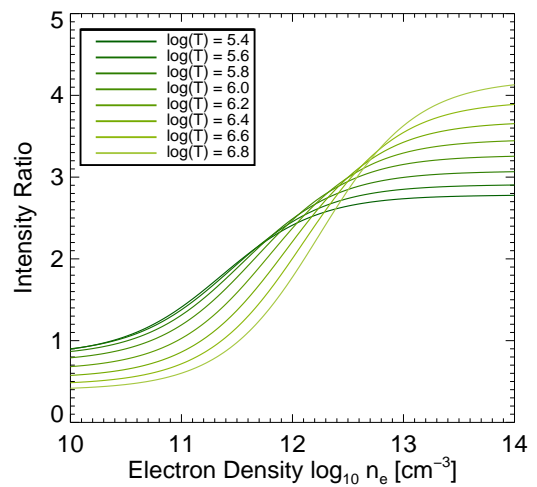}
  \caption{O~{\sc v} $\lambda$192/248 diagnostic curves calculated for varying plasma temperature.}
  \label{fig:ioneq}%
\end{figure}

During the flare onset the footpoint plasma temperature may rise faster than local ionisation processes can occur. For example the peak formation temperature of O~{\sc v} will be shifted to higher temperatures, although the modelling by \cite{2009A&A...502..409B} predicts that the effect is small at high densities. Nevertheless, we show the effect on the diagnostic curve in Figure \ref{fig:ioneq} where the diagnostic ratio has been calculated at various plasma temperatures; from the line formation temperature in equilibrium, to the peak of the temperature distribution of a hot footpoint \citep{2013ApJ...767...83G}.
For ratios close to those measured, i.e $\sim 2.3$ in Flare 2 FP2, the diagnostic density will range from $\log n_e = 11.85 - 12.25$ depending on the degree of non-equilibrium ionisation, therefore for a sufficiently rapid temperature change the measured densities may be a lower limit.



\section{Flare Heating of Dense Plasma}

After detailed analysis of high resolution \emph{Hinode}/EIS spectra we can confirm the presence of footpoint densities up to $\log n_e \sim 11.9$ at a temperature of $2.5 \times 10^{5}$~K.

Here we will investigate the energy input required to produce the observed temperature and density in the same two semi-empirical atmospheric models as Section \ref{sec:assumptions}; the VAL-E bright network model and the F1 flare model. Heating of the O~{\sc v} emitting region by both accelerated particles, and thermal conduction will be considered in turn. The heating rates by accelerated particles are also calculated for each model using a modified, completely ionised density structure, which is a reasonable assumption as the appearance of highly ionised states of oxygen, iron, and calcium is clear evidence that the emitting region is ionised by rapid heating and collisions \citep{2005ApJ...630..573A}.

\begin{figure*}
  \centering
   \subfloat{{(a)}\includegraphics[width=8cm]{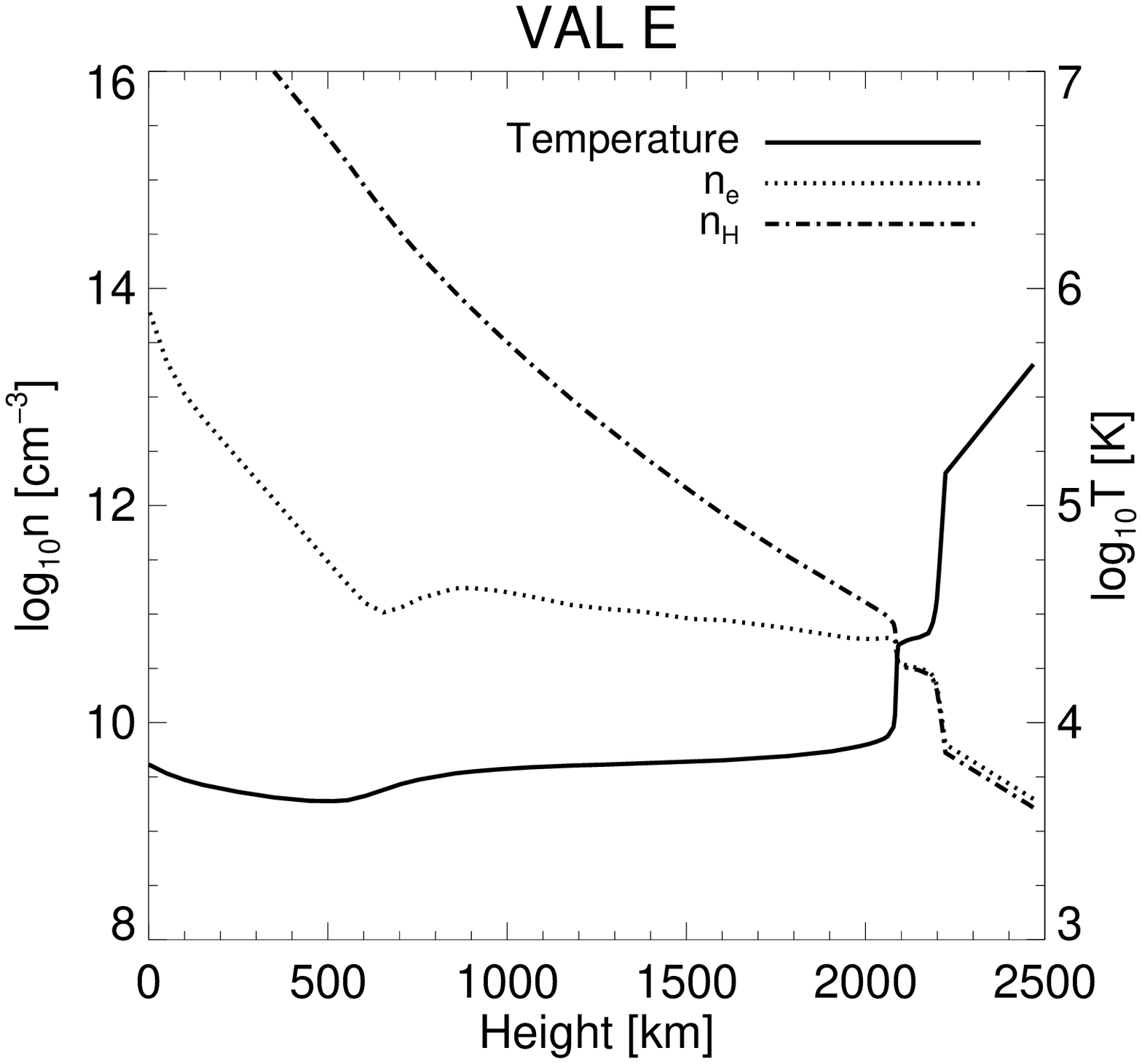}}
   \subfloat{{(b)}\includegraphics[width=8cm]{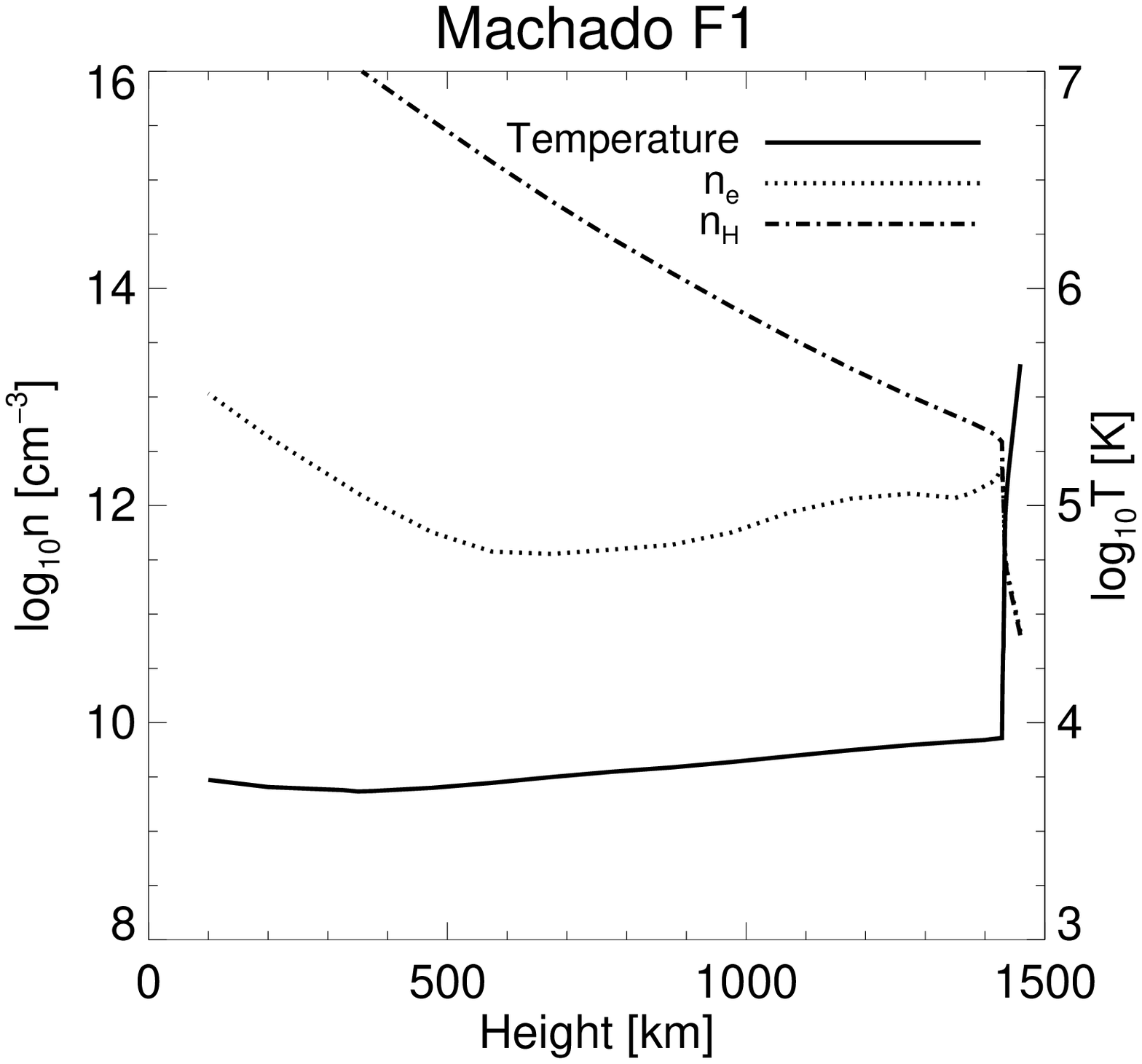}}
   
   \subfloat{{(c)}\includegraphics[width=8cm]{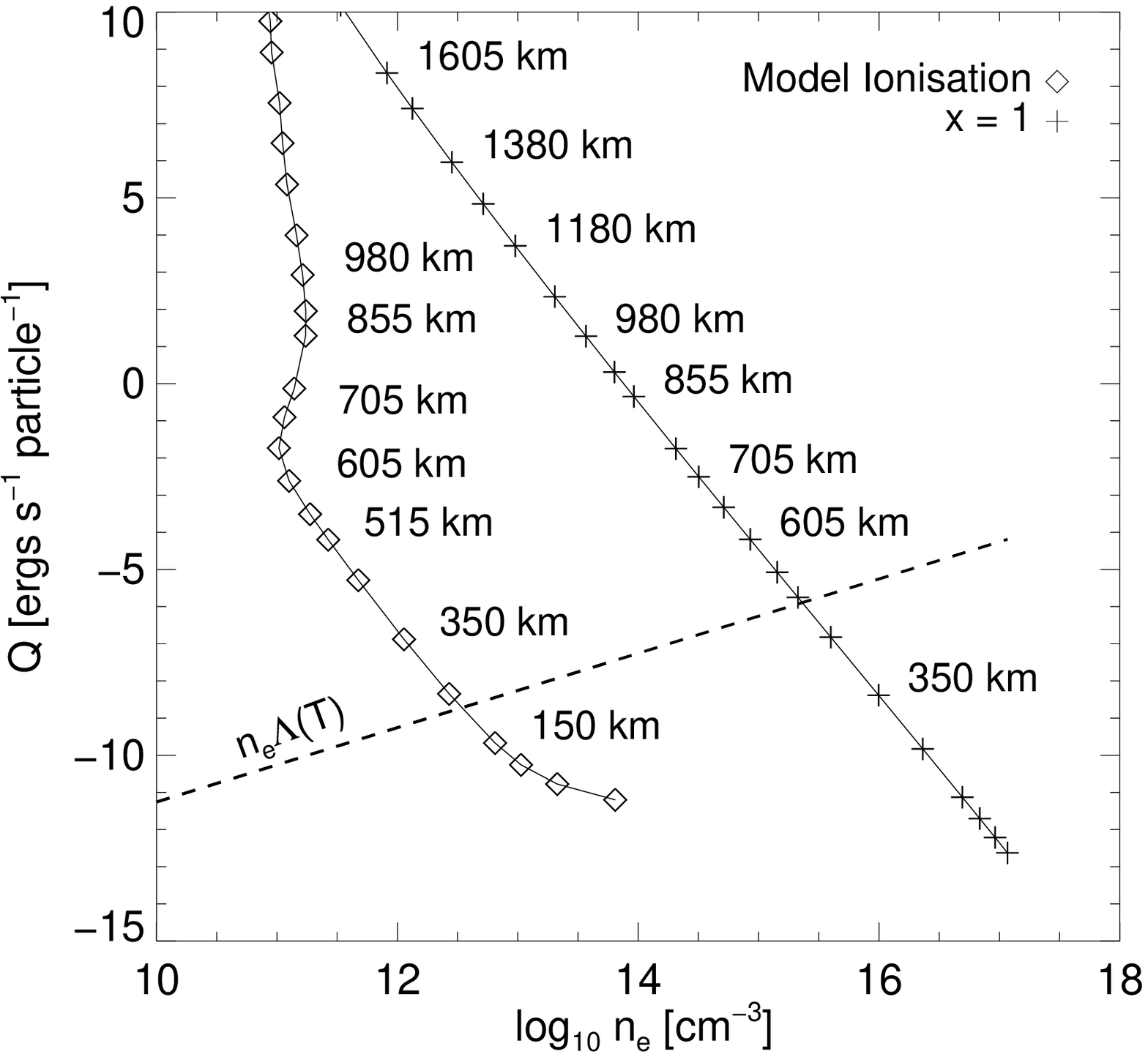}}
   \subfloat{{(d)}\includegraphics[width=8cm]{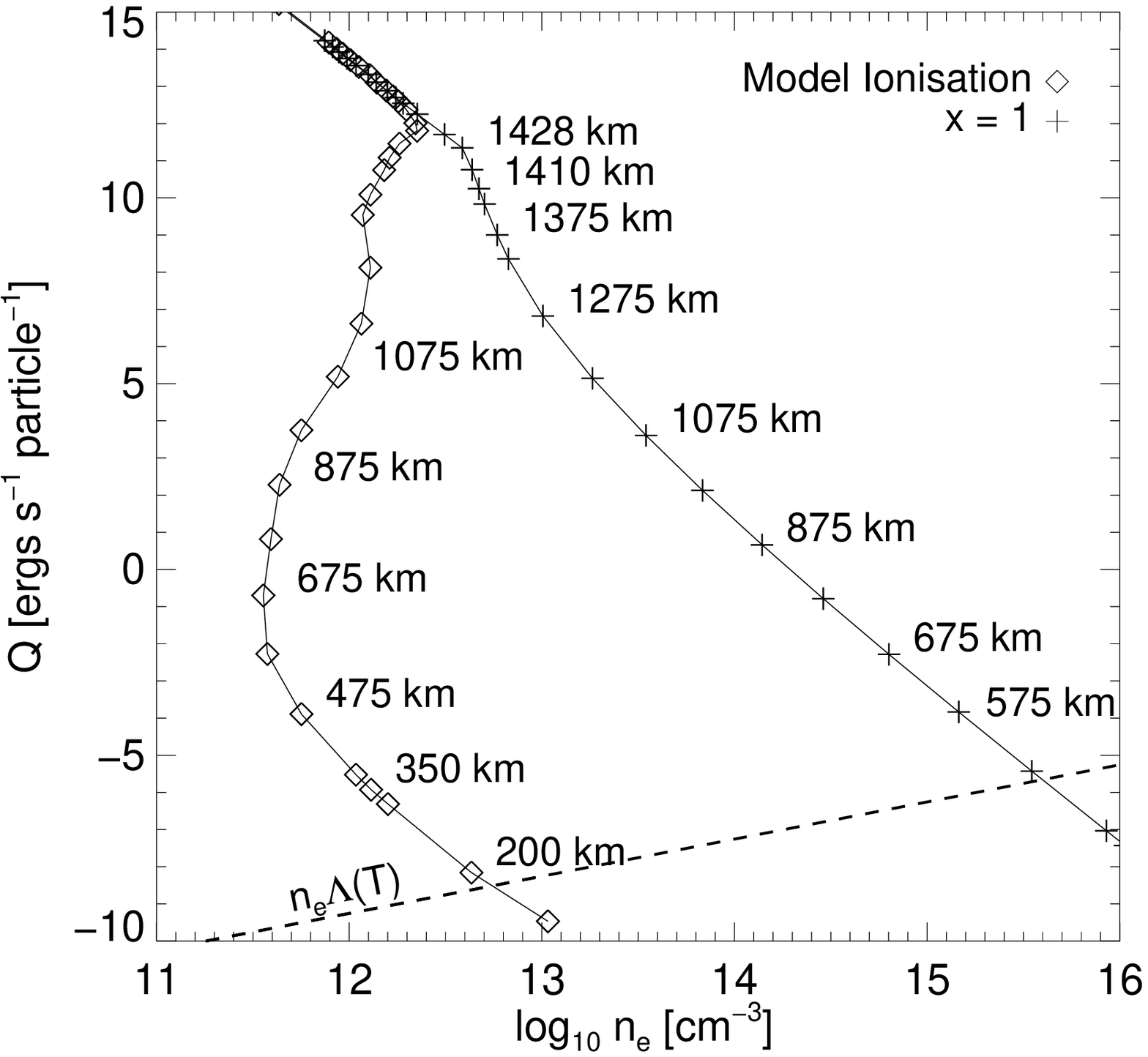}}

   \caption{Semi-empirical atmospheric models for a bright network element \citep[VAL-E]{1981ApJS...45..635V} and a flare \citep{1980ApJ...242..336M}, panels (a) \& (b), are plotted using tabulated results from their respective papers. In the lower panels, (c) \& (d), the heating rate per particle, deposited by streaming electrons in a collisional thick target regime, is shown as a function of density within the respective model atmosphere, and for both the model ionisation fraction (diamonds) and a modified atmosphere with constant ionisation fraction $x=1$ (crosses). The radiative loss rate per particle, $n_e \Lambda(T)$, at the O~{\sc v} formation temperature is also displayed.}
  \label{fig:models}%
\end{figure*}

The two model atmospheres are plotted in Figures \ref{fig:models} (a) and (b). In the VAL-E model, the electron density only reaches $\log n_e = 11.5$ below 500~km, well into the temperature minimum region. The structure of the F1 model is similar to VAL-E, except that the transition region has been `pushed' down to around 1400~km (note the different x-axis scales) and the density enhanced up to $\log n_e = 12$ at higher altitudes. The completely static model may be a strong assumption however, as pressure-driven density enhancements may occur during rapid heating. In future we plan to extend this work to include atmospheric structures calculated from radiative hydrodynamic codes, e.g \cite{1985ApJ...289..425F} and \cite{2005ApJ...630..573A}.

\subsection{Electron-beam heating of the O V emitting region}
The impulsive phase of Flare 2 (14:13~UT) was captured by RHESSI, and imaging and spectral analysis was presented in \cite{2009ApJ...699..968M}. A thermal plus non-thermal power law fit to the integrated spectra gave a low energy cut off of $E_C = 13 \pm 2~{\rm keV}$ with a spectral index of $\delta = 7.6 \pm 0.7$. The estimated area of $3 \times 10^{17}~{\rm cm}$ returned a flux in the non-thermal tail of $F_P = 5 \times 10^{10}~{\rm ergs~cm^{-2}~s^{-1}}$. In the following we use these parameters to calculate the heating rate per particle deposited by an injected distribution of non-thermal electrons, above the cut-off energy $E_C$, as a function of height within both model atmospheres. Flare 2 was the only event from this active region with consistent RHESSI coverage, although similar HXR parameters have been obtained in other medium sized events \citep{2011A&A...532A..27G}. We therefore expect this to be fairly typical among other small confined flares and representative also of Flare 1.

We calculate the heating rate per hydrogen nucleus as a function of column depth, $Q(N)$, for a collimated beam ($\mu_o = 1$) as given by \cite{1978ApJ...224..241E}, with corrections for $\beta$ \citep{1981ApJ...245..711E}. The rate is expressed as

\begin{equation}
  Q(N) = {1 \over 2} {K \gamma (\delta-2) \over \mu_o} B\left( {\delta \over 2}, {2\over (4 + \beta)} \right) {F_P \over {E_\mathrm{C}^2}} \left[{N} \over N_c \right]^{-\delta/2}~{\rm ergs~s^{-1}},
\end{equation}

where $K = 2 \pi e^4$,

\begin{equation}
\beta = {2 x \Lambda + (1 - x) \Lambda''} / {\Lambda' + x(\Lambda - \Lambda')},
\end{equation}

\begin{equation}
\gamma = x\Lambda + (1-x)\Lambda',
\end{equation}

and the Coulomb logarithm for an ionised target, $\Lambda$, and effective Coulomb logarithms for neutral targets, $\Lambda^{'}$ and $\Lambda^{''}$, are given in \cite{1994ApJ...426..387H}. The ionisation fraction, $x$, can be either assumed to be constant or estimated from the ratio of model electron density to hydrogen density.

The column depth, $N$, represents the quantity of material traversed by the electron beam. From the atmospheric profiles in Figures \ref{fig:models} (a) \& (b) we obtain the column depth as a function of height by integrating $\int_{z = s}^{z_{top}} n_H(S) dS $ at each position in the atmosphere, where $n_H$ is the hydrogen density and $S$ is the height in the model atmosphere. The heating rate, $Q$, is then easily evaluated at each height. To aid the interpretation of the diagnostic results, $Q$ is plotted in Figures \ref{fig:models} (c) and (d) against the model electron density, for both model derived and fully ionised density structures (diamonds and crosses respectively). As $Q$ is not expressed as a function of electron density, only of column depth, the reader should be aware that the heating rate can have two values at a single density. The model heights at some intermediate points are plotted for clarity.

Predictably, we find that the heating rate falls off rapidly with increasing depth, especially in the final few 100~km towards the photosphere. The optically thin radiative loss rate per particle, $n_e \Lambda(T)$, was taken from CHIANTI for the O~{\sc v} formation temperature above. The point where this line intersects the heating rate indicates where the atmosphere, if heated to O~{\sc v} temperatures, will radiate energy away as quickly as it is deposited, by optically-thin radiative losses. In deeper layers the energy deposition cannot deliver enough energy to raise the ambient temperature.

For the model ionisation structure, given by $n_e / n_H$, the critical density for zero net heating in the VAL-E model is approximately $\log n_e = 12.2$, and in the F1 atmosphere $\log n_e = 12.4$. The measurements for Flare 1 and 2 are below these values, putting them in a region which is possible to heat. The maximum density observed, $\log n_e = 11.89$, implies that heating by electrons in both atmospheres must reach a height of $\sim450$~km to produce the required O~{\sc v} emission. However, if the plasma is far from ionisation equilibrium (see Figure \ref{fig:ioneq}) the measured densities imply that the emitting region may be much closer to the heating limit.

If we assume that both atmospheres are completely ionised, i.e $n_e = n_H$, then the heating rate intersects the radiative loss rate at a density $\sim~3$ orders of magnitude higher. The O~{\sc v} emitting region therefore occurs much higher, at approximately 1600~km for VAL-E and 1450~km for F1, where the heating rate far exceeds the radiative losses. Without a further diagnostic to directly probe the ionisation state throughout the atmosphere we can only deduce that the location of heating lies somewhere between the two cases shown, i.e 450-1600~km.


 

\subsection{Conductive heating of the O V emitting region}

In addition to any source of direct heating there may be a significant or dominant conductive energy transport into the deeper layers of the atmosphere. The emission measure distributions obtained in \cite{2013ApJ...767...83G} were consistent with a flaring chromosphere that was in conductive and radiative balance, where a directly heated layer at 10~MK supplied a conductive flux to the chromosphere below, balanced by radiative losses.

In a steady-state atmosphere with no dynamic flows the energy balance equation is,

\begin{equation}
\nabla \cdot F_{\rm C} - E_{\rm R} + E_{\rm H}   = 0
\end{equation} 

where $F_C$ is the conductive flux, $E_R$ is the radiative loss rate, and $E_H$ the heating rate from additional sources. If we assume that no direct heating reaches the region where conductive energy input is balanced by radiative losses, then

\begin{equation}\label{eq:c2}
\nabla \cdot F_{\rm C} -n_{\rm e}^2 \Lambda(T) = 0
\end{equation} 

for a radiative loss rate $\Lambda(T)$ at the temperature of O~{\sc v} formation ($\log T = 5.4$). The conductive flux in 1-D is expressed as

\begin{equation}\label{eq:c3}
F_{\rm c} = \kappa_oT^{5/2}\frac{dT}{dS}
\end{equation}

where $\kappa_o = 9.2\times 10^{-7}$ for classical Spitzer conductivity. The temperature gradient or length scales are not known in the chromosphere, and the conductive flux can only be estimated. However, by making some simple assumptions we can estimate the scale of the emitting region required to balance a purely conductive heating term. In 1-D the conductive term in Equation \ref{eq:c2} then becomes

\begin{equation}
\frac{dF_{\rm c}}{dS} = \kappa_o\biggl(\frac{5}{2}T^{3/2}\frac{dT}{dS}\cdot\frac{dT}{dS} + T^{5/2}\frac{d^2T}{dS^2} \biggr).
\end{equation}

If the top of the chromosphere is impulsively heated to $T_{\rm max}=10$~MK then we can replace $dT/dS$ with $(T_{\rm max}~-~T)/L$, where $T$ is the temperature of the plasma that is emitting oxygen lines, and $L$ is the chromospheric scale length required to balance conductive energy inputs with radiative losses. By assuming a uniform temperature gradient, i.e $d^2T/dS^2 = 0$, the conductive flux for a scale length in units of km becomes

\begin{equation}
\frac{dF_{\rm c}}{dS}\sim \frac{2.75 \times 10^{6}}{L^2}.
\end{equation}

By rearranging Equation \ref{eq:c2}, and inserting the result above, the chromospheric length scale $L$ is

\begin{equation}
L = \left( {2.75 \times 10^{6} \over {n_e^2 \Lambda(T)}} \right)^{1/2}.
\end{equation}


Our requirement is to supply enough conductive flux to the chromosphere below a 10~MK layer to allow plasma at $T=10^{5.4}$~K to radiate. The density of the chromosphere will vary with depth, but if we assume that the average density of the whole region is between $10^{11}-10^{12}~{\rm cm^{-3}}$ then the conductive input will be balanced by the radiative losses for a chromosphere of thickness 70-700~km. Inspecting the atmospheres in Figures \ref{fig:models} (a) and (b) shows that these thicknesses are reasonable within the picture of a compressed, lower altitude transition region. However, the semi-empirical models shown do not include a sufficiently high temperature component at $\sim$10~MK, as found by several more recent studies \citep{2004A&A...415..377M, 2012ApJ...754...54B, 2013ApJ...767...83G, 2013ApJ...771..104F}.

\section{Conclusions}\label{sec:conclusions}

The location, or depth in a 1-D interpretation, of energy deposition within the flaring chromosphere remains an important constraint on the global plasma behaviour of the flaring chromosphere. We have presented in this paper new confirmation of 250,000~K footpoint plasma with electron densities of $\log n_e~({\rm cm^{-3}}) = 11.3 - 11.9$ in two flares. The assumption of optically-thin plasma was tested and found to be acceptable for these lines using two model atmospheres. The effect of non-equilibrium ionisation on the diagnostic was taken into consideration and found to alter the measured density, potentially raising the maximum density observed to $\log n_e~({\rm cm^{-3}}) \sim 12.2$, although further modelling is required to determine how significant the effect is during the impulsive phase.


With our measurement of the electron density (and knowledge of the temperature) of the O~{\sc v} emitting region we have been able to estimate its properties under different assumptions about the energy input. Assuming an atmosphere structured according to VAL-E or F1, the emitting region is found $\sim$450~km above the photosphere. The collisional beam heating rate is found to exceed the radiative losses here, although if the density was slightly higher, $\log n_e~({\rm cm^{-3}}) > 12.4$, then radiative losses would balance the heating input.

If the calculation is repeated for the VAL-E and F1 mass density structure but assuming complete ionisation then the O~{\sc v} emitting region is found around 1450-1600~km above the photosphere, where the heating input rate easily exceeds the radiative losses. As the ionisation fraction is unknown we determine the heating location to be between 450-1600~km.

Energy transport from a directly heated slab via a conduction flux was also considered and can also provide the necessary O~{\sc v} emission given a chromosphere of thickness 70-700~km, depending on the model density, which is consistent with the emission depths that we have estimated. A combination of both transport methods will be considered in future, including conductive losses to the lower atmosphere.

The ionisation state of the flaring chromosphere clearly plays an important role in determining the energy balance in flare footpoints and on diagnostics of this kind. By coupling further temperature and density diagnostics with radiative-hydrodynamic models, and including new observations of ionisation/recombination signatures \citep{2014ApJ...794L..23H}, then the location of the emitting region may be better constrained.





Finally, this work shows that diagnostics of high density plasmas are of great value, as was seen in the past by the \emph{Skylab} and UVSP spectrographs, yet few studies of this kind have been made in recent years. Whilst it is possible with EIS to make this particular diagnostic, the long wavelength detector is degrading \citep{2013A&A...555A..47D} and the 248\AA\ line is becoming difficult to observe without a specifically targeted EIS study (longer exposures may be necessary which are not best suited to flare observations). The Extreme-Ultraviolet Variability Experiment \citep{2012SoPh..275..115W} measures the strong O~{\sc v}~629.73\AA\ to 760.30\AA\ ratio among others, but footpoint identification is difficult with no spatial resolution. Data from IRIS is extremely promising though, as the O~{\sc iv} 1404.78\AA\ to 1399.77\AA\ ratio \citep{2014ApJ...780L..12D} should provide a useful diagnostic at transition region temperatures and with improved spatial and temporal resolution.


\begin{acknowledgements}
DRG acknowledges support from an STFC-STEP award. LF and NL acknowledge support from STFC-funded rolling grant ST/F002637/1. The research leading to these results has received funding from the European Community’s Seventh Framework Programme (FP7/2007-2013) under grant agreement no. 606862 (F-CHROMA). CHIANTI is a collaborative project involving the NRL (USA), the Universities of Florence (Italy) and Cambridge (UK), and George Mason University (USA). We are grateful for the open data policies of RHESSI and {\it Hinode} and the efforts of the instrument and software teams. Hinode is a Japanese mission developed and launched by ISAS/JAXA, with NAOJ as domestic partner and NASA and STFC (UK) as international partners. It is operated by these agencies in co-operation with ESA and NSC (Norway). We would also like to thank Peter Young and Jaroslav Dudik for their discussion and helpful comments.
\end{acknowledgements}

\bibliographystyle{aa}
\bibliography{thesis}

\begin{thebibliography}{52}
\expandafter\ifx\csname natexlab\endcsname\relax\def\natexlab#1{#1}\fi

\bibitem[{{Allred} {et~al.}(2005){Allred}, {Hawley}, {Abbett}, \&
  {Carlsson}}]{2005ApJ...630..573A}
{Allred}, J.~C., {Hawley}, S.~L., {Abbett}, W.~P., \& {Carlsson}, M. 2005,
  \apj, 630, 573

\bibitem[{{Antonucci} \& {Dennis}(1983)}]{1983SoPh...86...67A}
{Antonucci}, E. \& {Dennis}, B.~R. 1983, \solphys, 86, 67

\bibitem[{{Bartoe} {et~al.}(1977){Bartoe}, {Brueckner}, {Purcell}, \&
  {Tousey}}]{1977ApOpt..16..879B}
{Bartoe}, J.-D.~F., {Brueckner}, G.~E., {Purcell}, J.~D., \& {Tousey}, R. 1977,
  \ao, 16, 879

\bibitem[{{Bradshaw}(2009)}]{2009A&A...502..409B}
{Bradshaw}, S.~J. 2009, \aap, 502, 409

\bibitem[{{Brosius}(2012)}]{2012ApJ...754...54B}
{Brosius}, J.~W. 2012, \apj, 754, 54

\bibitem[{{Brown} {et~al.}(2009){Brown}, {Turkmani}, {Kontar}, {MacKinnon}, \&
  {Vlahos}}]{2009A&A...508..993B}
{Brown}, J.~C., {Turkmani}, R., {Kontar}, E.~P., {MacKinnon}, A.~L., \&
  {Vlahos}, L. 2009, \aap, 508, 993

\bibitem[{{Cheng} {et~al.}(1982){Cheng}, {Bruner}, {Tandberg-Hanssen},
  {Woodgate}, {Shine}, {Kenny}, {Henze}, \& {Poletto}}]{1982ApJ...253..353C}
{Cheng}, C.-C., {Bruner}, E.~C., {Tandberg-Hanssen}, E., {et~al.} 1982, \apj,
  253, 353

\bibitem[{{Czaykowska} {et~al.}(1999){Czaykowska}, {De Pontieu}, {Alexander},
  \& {Rank}}]{1999ApJ...521L..75C}
{Czaykowska}, A., {De Pontieu}, B., {Alexander}, D., \& {Rank}, G. 1999, \apjl,
  521, L75

\bibitem[{{De Pontieu} {et~al.}(2014){De Pontieu}, {Title}, {Lemen}, {Kushner},
  {Akin}, {Allard}, {Berger}, {Boerner}, {Cheung}, {Chou}, {Drake}, {Duncan},
  {Freeland}, {Heyman}, {Hoffman}, {Hurlburt}, {Lindgren}, {Mathur}, {Rehse},
  {Sabolish}, {Seguin}, {Schrijver}, {Tarbell}, {W{\"u}lser}, {Wolfson},
  {Yanari}, {Mudge}, {Nguyen-Phuc}, {Timmons}, {van Bezooijen}, {Weingrod},
  {Brookner}, {Butcher}, {Dougherty}, {Eder}, {Knagenhjelm}, {Larsen},
  {Mansir}, {Phan}, {Boyle}, {Cheimets}, {DeLuca}, {Golub}, {Gates}, {Hertz},
  {McKillop}, {Park}, {Perry}, {Podgorski}, {Reeves}, {Saar}, {Testa}, {Tian},
  {Weber}, {Dunn}, {Eccles}, {Jaeggli}, {Kankelborg}, {Mashburn}, {Pust},
  {Springer}, {Carvalho}, {Kleint}, {Marmie}, {Mazmanian}, {Pereira}, {Sawyer},
  {Strong}, {Worden}, {Carlsson}, {Hansteen}, {Leenaarts}, {Wiesmann},
  {Aloise}, {Chu}, {Bush}, {Scherrer}, {Brekke}, {Martinez-Sykora}, {Lites},
  {McIntosh}, {Uitenbroek}, {Okamoto}, {Gummin}, {Auker}, {Jerram}, {Pool}, \&
  {Waltham}}]{2014SoPh..289.2733D}
{De Pontieu}, B., {Title}, A.~M., {Lemen}, J.~R., {et~al.} 2014, \solphys, 289,
  2733

\bibitem[{{Del Zanna}(2013)}]{2013A&A...555A..47D}
{Del Zanna}, G. 2013, \aap, 555, A47

\bibitem[{{Del Zanna} {et~al.}(2011){Del Zanna}, {Mitra-Kraev}, {Bradshaw},
  {Mason}, \& {Asai}}]{2011A&A...526A...1D}
{Del Zanna}, G., {Mitra-Kraev}, U., {Bradshaw}, S.~J., {Mason}, H.~E., \&
  {Asai}, A. 2011, \aap, 526, A1

\bibitem[{{Dere} {et~al.}(1997){Dere}, {Landi}, {Mason}, {Monsignori Fossi}, \&
  {Young}}]{1997A&AS..125..149D}
{Dere}, K.~P., {Landi}, E., {Mason}, H.~E., {Monsignori Fossi}, B.~C., \&
  {Young}, P.~R. 1997, \aaps, 125, 149

\bibitem[{{Doschek} {et~al.}(1977){Doschek}, {Feldman}, \&
  {Rosenberg}}]{1977ApJ...215..329D}
{Doschek}, G.~A., {Feldman}, U., \& {Rosenberg}, F.~D. 1977, \apj, 215, 329

\bibitem[{{Dud{\'{\i}}k} {et~al.}(2014){Dud{\'{\i}}k}, {Del Zanna}, {Dzif{\v
  c}{\'a}kov{\'a}}, {Mason}, \& {Golub}}]{2014ApJ...780L..12D}
{Dud{\'{\i}}k}, J., {Del Zanna}, G., {Dzif{\v c}{\'a}kov{\'a}}, E., {Mason},
  H.~E., \& {Golub}, L. 2014, \apjl, 780, L12

\bibitem[{{Dzif{\v c}{\'a}kov{\'a}} \&
  {Kulinov{\'a}}(2011)}]{2011A&A...531A.122D}
{Dzif{\v c}{\'a}kov{\'a}}, E. \& {Kulinov{\'a}}, A. 2011, \aap, 531, A122

\bibitem[{{Emslie}(1978)}]{1978ApJ...224..241E}
{Emslie}, A.~G. 1978, \apj, 224, 241

\bibitem[{{Emslie}(1981)}]{1981ApJ...245..711E}
{Emslie}, A.~G. 1981, \apj, 245, 711

\bibitem[{{Feldman} {et~al.}(1977){Feldman}, {Dorschek}, \&
  {Rosenberg}}]{1977ApJ...215..652F}
{Feldman}, U., {Dorschek}, G.~A., \& {Rosenberg}, F.~D. 1977, \apj, 215, 652

\bibitem[{{Fisher} {et~al.}(1985){Fisher}, {Canfield}, \&
  {McClymont}}]{1985ApJ...289..425F}
{Fisher}, G.~H., {Canfield}, R.~C., \& {McClymont}, A.~N. 1985, \apj, 289, 425

\bibitem[{{Fletcher} {et~al.}(2013){Fletcher}, {Hannah}, {Hudson}, \&
  {Innes}}]{2013ApJ...771..104F}
{Fletcher}, L., {Hannah}, I.~G., {Hudson}, H.~S., \& {Innes}, D.~E. 2013, \apj,
  771, 104

\bibitem[{{Fletcher} \& {Hudson}(2008)}]{2008ApJ...675.1645F}
{Fletcher}, L. \& {Hudson}, H.~S. 2008, \apj, 675, 1645

\bibitem[{{Graham} {et~al.}(2011){Graham}, {Fletcher}, \&
  {Hannah}}]{2011A&A...532A..27G}
{Graham}, D.~R., {Fletcher}, L., \& {Hannah}, I.~G. 2011, \aap, 532, A27

\bibitem[{{Graham} {et~al.}(2013){Graham}, {Hannah}, {Fletcher}, \&
  {Milligan}}]{2013ApJ...767...83G}
{Graham}, D.~R., {Hannah}, I.~G., {Fletcher}, L., \& {Milligan}, R.~O. 2013,
  \apj, 767, 83

\bibitem[{{Grevesse} \& {Sauval}(1998)}]{1998SSRv...85..161G}
{Grevesse}, N. \& {Sauval}, A.~J. 1998, \ssr, 85, 161

\bibitem[{{Hawley} \& {Fisher}(1994)}]{1994ApJ...426..387H}
{Hawley}, S.~L. \& {Fisher}, G.~H. 1994, \apj, 426, 387

\bibitem[{{Heinzel} \& {Kleint}(2014)}]{2014ApJ...794L..23H}
{Heinzel}, P. \& {Kleint}, L. 2014, \apjl, 794, L23

\bibitem[{{Keenan} {et~al.}(2014){Keenan}, {Doyle}, {Madjarska}, {Rose},
  {Bowler}, {Britton}, {McCrink}, \& {Mathioudakis}}]{2014ApJ...784L..39K}
{Keenan}, F.~P., {Doyle}, J.~G., {Madjarska}, M.~S., {et~al.} 2014, \apjl, 784,
  L39

\bibitem[{{Keenan} {et~al.}(1991){Keenan}, {Dufton}, {Harra}, {Conlon},
  {Berrington}, {Kingston}, \& {Widing}}]{1991ApJ...382..349K}
{Keenan}, F.~P., {Dufton}, P.~L., {Harra}, L.~K., {et~al.} 1991, \apj, 382, 349

\bibitem[{{Kerr} {et~al.}(2004){Kerr}, {Rose}, {Wark}, \&
  {Keenan}}]{2004ApJ...613L.181K}
{Kerr}, F.~M., {Rose}, S.~J., {Wark}, J.~S., \& {Keenan}, F.~P. 2004, \apjl,
  613, L181

\bibitem[{{Ko} {et~al.}(2009){Ko}, {Doschek}, {Warren}, \&
  {Young}}]{2009ApJ...697.1956K}
{Ko}, Y.-K., {Doschek}, G.~A., {Warren}, H.~P., \& {Young}, P.~R. 2009, \apj,
  697, 1956

\bibitem[{{Kulinov{\'a}} {et~al.}(2011){Kulinov{\'a}}, {Ka{\v s}parov{\'a}},
  {Dzif{\v c}{\'a}kov{\'a}}, {Sylwester}, {Sylwester}, \&
  {Karlick{\'y}}}]{2011A&A...533A..81K}
{Kulinov{\'a}}, A., {Ka{\v s}parov{\'a}}, J., {Dzif{\v c}{\'a}kov{\'a}}, E.,
  {et~al.} 2011, \aap, 533, A81

\bibitem[{{Landi} {et~al.}(2013){Landi}, {Young}, {Dere}, {Del Zanna}, \&
  {Mason}}]{2013ApJ...763...86L}
{Landi}, E., {Young}, P.~R., {Dere}, K.~P., {Del Zanna}, G., \& {Mason}, H.~E.
  2013, \apj, 763, 86

\bibitem[{{Lin} {et~al.}(2002){Lin}, {Dennis}, {Hurford}, {Smith}, {Zehnder},
  {Harvey}, {Curtis}, {Pankow}, {Turin}, {Bester}, {Csillaghy}, {Lewis},
  {Madden}, {van Beek}, {Appleby}, {Raudorf}, {McTiernan}, {Ramaty}, {Schmahl},
  {Schwartz}, {Krucker}, {Abiad}, {Quinn}, {Berg}, {Hashii}, {Sterling},
  {Jackson}, {Pratt}, {Campbell}, {Malone}, {Landis}, {Barrington-Leigh},
  {Slassi-Sennou}, {Cork}, {Clark}, {Amato}, {Orwig}, {Boyle}, {Banks},
  {Shirey}, {Tolbert}, {Zarro}, {Snow}, {Thomsen}, {Henneck}, {McHedlishvili},
  {Ming}, {Fivian}, {Jordan}, {Wanner}, {Crubb}, {Preble}, {Matranga}, {Benz},
  {Hudson}, {Canfield}, {Holman}, {Crannell}, {Kosugi}, {Emslie}, {Vilmer},
  {Brown}, {Johns-Krull}, {Aschwanden}, {Metcalf}, \&
  {Conway}}]{2002SoPh..210....3L}
{Lin}, R.~P., {Dennis}, B.~R., {Hurford}, G.~J., {et~al.} 2002, \solphys, 210,
  3

\bibitem[{{Liu} {et~al.}(2009){Liu}, {Petrosian}, \&
  {Mariska}}]{2009ApJ...702.1553L}
{Liu}, W., {Petrosian}, V., \& {Mariska}, J.~T. 2009, \apj, 702, 1553

\bibitem[{{Machado} {et~al.}(1980){Machado}, {Avrett}, {Vernazza}, \&
  {Noyes}}]{1980ApJ...242..336M}
{Machado}, M.~E., {Avrett}, E.~H., {Vernazza}, J.~E., \& {Noyes}, R.~W. 1980,
  \apj, 242, 336

\bibitem[{{Mariska} \& {Poland}(1985)}]{1985SoPh...96..317M}
{Mariska}, J.~T. \& {Poland}, A.~I. 1985, \solphys, 96, 317

\bibitem[{{Mihalas}(1978)}]{1978stat.book.....M}
{Mihalas}, D. 1978, {Stellar atmospheres /2nd edition/}

\bibitem[{{Milligan}(2011)}]{2011ApJ...740...70M}
{Milligan}, R.~O. 2011, \apj, 740, 70

\bibitem[{{Milligan} \& {Dennis}(2009)}]{2009ApJ...699..968M}
{Milligan}, R.~O. \& {Dennis}, B.~R. 2009, \apj, 699, 968

\bibitem[{{Mrozek} \& {Tomczak}(2004)}]{2004A&A...415..377M}
{Mrozek}, T. \& {Tomczak}, M. 2004, \aap, 415, 377

\bibitem[{{Nagai} \& {Emslie}(1984)}]{1984ApJ...279..896N}
{Nagai}, F. \& {Emslie}, A.~G. 1984, \apj, 279, 896

\bibitem[{{Poland} {et~al.}(1984){Poland}, {Orwig}, {Mariska}, {Auer}, \&
  {Nakatsuka}}]{1984ApJ...280..457P}
{Poland}, A.~I., {Orwig}, L.~E., {Mariska}, J.~T., {Auer}, L.~H., \&
  {Nakatsuka}, R. 1984, \apj, 280, 457

\bibitem[{{Russell} \& {Fletcher}(2013)}]{2013ApJ...765...81R}
{Russell}, A.~J.~B. \& {Fletcher}, L. 2013, \apj, 765, 81

\bibitem[{{Vernazza} {et~al.}(1981){Vernazza}, {Avrett}, \&
  {Loeser}}]{1981ApJS...45..635V}
{Vernazza}, J.~E., {Avrett}, E.~H., \& {Loeser}, R. 1981, \apjs, 45, 635

\bibitem[{{Warren} {et~al.}(2014){Warren}, {Ugarte-Urra}, \&
  {Landi}}]{2014ApJS..213...11W}
{Warren}, H.~P., {Ugarte-Urra}, I., \& {Landi}, E. 2014, \apjs, 213, 11

\bibitem[{{Watanabe} {et~al.}(2010){Watanabe}, {Hara}, {Sterling}, \&
  {Harra}}]{2010ApJ...719..213W}
{Watanabe}, T., {Hara}, H., {Sterling}, A.~C., \& {Harra}, L.~K. 2010, \apj,
  719, 213

\bibitem[{{Widing} {et~al.}(1982){Widing}, {Doyle}, {Dufton}, \&
  {Kingston}}]{1982ApJ...257..913W}
{Widing}, K.~G., {Doyle}, J.~G., {Dufton}, P.~L., \& {Kingston}, E.~A. 1982,
  \apj, 257, 913

\bibitem[{{Woodgate} {et~al.}(1980){Woodgate}, {Brandt}, {Kalet}, {Kenny},
  {Tandberg-Hanssen}, {Bruner}, {Beckers}, {Henze}, {Knox}, \&
  {Hyder}}]{1980SoPh...65...73W}
{Woodgate}, B.~E., {Brandt}, J.~C., {Kalet}, M.~W., {et~al.} 1980, \solphys,
  65, 73

\bibitem[{{Woods} {et~al.}(2012){Woods}, {Eparvier}, {Hock}, {Jones},
  {Woodraska}, {Judge}, {Didkovsky}, {Lean}, {Mariska}, {Warren}, {McMullin},
  {Chamberlin}, {Berthiaume}, {Bailey}, {Fuller-Rowell}, {Sojka}, {Tobiska}, \&
  {Viereck}}]{2012SoPh..275..115W}
{Woods}, T.~N., {Eparvier}, F.~G., {Hock}, R., {et~al.} 2012, \solphys, 275,
  115

\bibitem[{{Young} {et~al.}(2007){Young}, {Del Zanna}, {Mason}, {Doschek},
  {Culhane}, \& {Hara}}]{2007PASJ...59S.727Y}
{Young}, P.~R., {Del Zanna}, G., {Mason}, H.~E., {et~al.} 2007, \pasj, 59, 727

\bibitem[{{Young} {et~al.}(2015){Young}, {Tian}, \&
  {Jaeggli}}]{2015ApJ...799..218Y}
{Young}, P.~R., {Tian}, H., \& {Jaeggli}, S. 2015, \apj, 799, 218

\bibitem[{{Young} {et~al.}(2009){Young}, {Watanabe}, {Hara}, \&
  {Mariska}}]{2009A&A...495..587Y}
{Young}, P.~R., {Watanabe}, T., {Hara}, H., \& {Mariska}, J.~T. 2009, \aap,
  495, 587

\end{thebibliography}
\end{document}